\documentclass{IEEEtran}
\usepackage{cite}
\usepackage{amsmath,amssymb,amsfonts}
\usepackage{algorithmic}
\usepackage{graphicx}
\usepackage{textcomp}
\usepackage{tikz}
\usepackage{quantikz}
\usepackage{subfig}
\usepackage{subfloat}
\usepackage{caption}
\usepackage{orcidlink}
\usepackage{pdfpages}

\newcommand{\ignore}[1]{}

\date{}

\begin{document}


\title{Quantum Circuit-Based Adaptation for Credit Risk Analysis}

\author{\uppercase{Halima Giovanna Ahmad}${}^{1}$, \IEEEmembership{Member,IEEE}\orcidlink{0000-0003-2627-2496}, \uppercase{Alessandro Sarno}${}^{1}$\orcidlink{0009-0005-1912-7540},  \uppercase{Mehdi El Bakraoui}${}^{2}$\orcidlink{0009-0002-4775-7943}, Carlo Cosenza${}^{1}$\orcidlink{0009-0004-2426-4008}, Clément Bésoin${}^{2}$\orcidlink{0009-0009-8141-7003}, Francesca Cibrario${}^{3}$\orcidlink{0009-0007-8290-4992}, Valeria Zaffaroni${}^{3}$\orcidlink{0009-0002-8221-5904}, Giacomo Ranieri${}^{3}$\orcidlink{0009-0005-0488-2121}, Roberto Bertilone${}^{3}$\orcidlink{0009-0001-1553-0835}, Viviana Stasino${}^{1}$\orcidlink{0009-0009-4898-5834}, Pasquale Mastrovito${}^{1}$\orcidlink{0000-0002-7833-0398}, Francesco Tafuri${}^{1}$\orcidlink{0000-0003-0784-1454}, Davide Massarotti${}^{4}$\orcidlink{0000-0001-7495-362X}, Leonardo Chabbra${}^{2}$\orcidlink{0009-0003-8839-041X}, Davide Corbelletto${}^{3}$\orcidlink{0009-0003-8830-2619}\\
	${}^{1}$Dipartimento di Fisica "Ettore Pancini", Università degli Studi di Napoli "Federico II", Via Cinthia, 80126, Napoli, IT.\\${}^{2}$G2Q Computing, Via A. Bertani, 2, Milano, IT.\\${}^{3}$Intesa Sanpaolo, Torino, IT.\\${}^{4}$Dipartimento di Ingegneria Elettrica e delle Tecnologie dell’Informazione, Università degli Studi di Napoli Federico II, Via Claudio 21, 80125, Napoli, IT. Corresponding author: Halima Giovanna Ahmad (email: halimagiovanna.ahmad\@ unina.it).
}




\maketitle

\section*{}

\textbf{Noisy and Intermediate-Scale Quantum, or NISQ, processors are sensitive to noise, prone to quantum decoherence, and are not yet capable of continuous quantum error correction for fault-tolerant quantum computation. Hence, quantum algorithms designed in the pre-fault-tolerant era cannot neglect the noisy nature of the hardware, and investigating the relationship between quantum hardware performance and the output of quantum algorithms is essential. In this work, we experimentally study how hardware-aware variational quantum circuits on a superconducting quantum processing unit can model distributions relevant to specific use-case applications for Credit Risk Analysis, e.g., standard Gaussian distributions for latent factor loading in the Gaussian Conditional-Independence model. We use a transpilation technique tailored to the specific quantum hardware topology, which minimizes gate depth and connectivity violations, and we calibrate the gate rotations of the circuit to achieve an optimized output from quantum algorithms. Our results demonstrate the viability of quantum adaptation on a small-scale, proof-of-concept model inspired by financial applications and offer a good starting point for understanding the practical use of NISQ devices. \footnote{This work has been accepted by IEEE. Copyright may be transferred without notice, after which this version may no longer be available.}}


\section{\label{sec:level1}Introduction}

Quantum computing has been promised as a reliable solution for use cases that are inefficiently addressed by standard classical computing. Among them, financial applications have attracted considerable attention lately. In the Credit Risk Analysis (CRA) domain, quantum strategies~\cite{Egger2021,Dri2023,veronelli2025implementingcreditriskanalysis} for computing quantities related to the estimation of Economic Capital, such as the Value at Risk (VaR) and the Conditional Value at Risk (CVaR), rely on Quantum Amplitude Estimation (QAE). This technique promises a theoretical quadratic speed-up with respect to its classical counterpart, the Monte Carlo simulation approach~\cite{Montanaro2015}.
However, the reduced number of addressable qubits and the relatively high level of gate errors and decoherence in state-of-the-art Quantum Processing Units (QPUs) represent a technological challenge to overcome for the demonstration of quantum algorithms, at least as proof-of-concepts~\cite{Chen2023}. Nevertheless, great advances in hardware performance enable us to explore several use cases with remarkable results already in the Noisy and Intermediate-Scale Quantum (NISQ) era~\cite{Chen2023}. Especially relevant in this framework are superconducting QPUs, which have attracted noticeable interest worldwide for their remarkable circuit analogies with CMOS-based classical processors, sharing similar historical bottlenecks, advantages, limitations, perspectives, and compatibilities with state-of-the-art classical computing techniques~\cite{Devoret2013}. Remarkably, superconducting technology provides extensive engineering freedom, which allows for an enriching day by day pool of applications in the quantum technologies realm~\cite{Salvoni2021,Mastrovito2024} and a strong variety of hardware platforms for quantum computing, each designed to tackle specific limitations of current quantum circuit designs~\cite{Rasmussen2021}. Superconducting QPUs (sQPUs) leverage the use of Josephson junctions to artificially engineer macroscopic artificial atoms in a way that allows easy preparation, manipulation, and readout of quantum states with an increasing level of complexity~\cite{Krantz2019,ahmad2023,stasino2025implementation}, and eventually exploit the fundamental computational resources of quantum computing: superposition and entanglement. Most importantly, the practical interface between sQPUs and classical computing platforms, e.g., cloud computing and High-Performance Computing (HPC), has provoked a strong speed-up and quantum awareness in the field of quantum algorithms~\cite{Beck2024,Mansfield2025}. Although pushing coherence times well above current limits and inventing smart ways to overcome gate and decoherence errors~\cite{Ahmad2022_ferro,Hu2024,Tuokkola2025} is probably the most sound solution to reach the fault-tolerant quantum (FTQ) regime~\cite{Preskill1998}, it is still worth systematically and coherently investigating the possible outcomes achievable in NISQ devices.  

In this work, we exploit a transmon-based sQPU, the core of the Italian Superconducting Quantum Computing Center \emph{Partenope}, conceived to experimentally address quantum algorithms directly at the hardware (or pulse) level. We report here the first experimental investigation and implementation of one sub-circuit in the QAE-based Credit Risk Analysis algorithm on this quantum machine. In this quantum procedure, the first part of the Grover operator’s state preparation is dedicated to loading an \emph{uncertainty model} that represents the probability of default of the counterparties under analysis. To perform preliminary experiments on quantum hardware, we decided to focus on this component since it remains unchanged when estimating both the VaR and the CVaR through repeated applications of the Grover operator. Moreover, the construction of the uncertainty model requires the loading of standard normal distributions $\mathcal{N}(0,1)$, with a mean equal to $0$ and a standard deviation of $1$, a task that is also reusable for other quantum finance use cases, such as option pricing~\cite{Stamatopoulos2020,chakrabarti2021,Cibrario2024,cibrario2025autocallableoptionspricingintegrationbased}. Concerning the execution of the \emph{uncertainty model}, the \emph{transpilation}, i.\,e. the process of converting a quantum circuit from one representation to another (often to optimize it for a specific quantum hardware platform) plays a key role, especially in the NISQ era and in QPUs characterized by limited connectivity (as the sQPU analyzed in this work). Akin to compilation in classical computing, transpilation adapts quantum circuits for the unique characteristics and limitations of different quantum computers instead of transforming code for different processors. Although in the long term quantum computing should be hardware agnostic, it is still not the case for NISQ devices. This work, in fact, aims to bridge a fundamental gap between quantum algorithm specialists and quantum hardware engineers, discussing in detail at the machine level the role of (i) hardware circuit parameters, (ii) qubit connectivity, (iii) hardware performance, (iv) electronics implementation of quantum logic gates on the actual output of a quantum algorithm, and the transpilation process that allows optimizing the outcome of a quantum circuit.  
\\
The paper is organized as follows. Sec.~\ref{sec:methods} defines how the uncertainty model is mapped to a quantum circuit. We present the scalable, parameterized circuit for a single-counterparty model, whose probability of default is influenced by a single risk-factor, and describe how each component of the circuit was developed. Sec.~\ref{sec:hardware} details the hardware we used for the executions — qubit topology and gate fidelities — and shows the transpiled circuit that reflects the device’s coupling map, which differs from the idealized layout. This allows us to detail the connection between the simulated transpilation of the quantum circuits and the effective transpilation required by the machine to achieve optimal results. Sec.~\ref{sec:results} presents results and a comparison of the quantum implementation with a classical baseline for the same model. Finally, in Sec.~\ref{sec:discussions} and Sec.~\ref{sec:conclusion}  we discuss on the scientific significance of this work, main advantages and limitations and on future perspectives.

\section{Methods}
\label{sec:methods}

\subsection{Context of Credit Risk Analysis and Gaussian Conditional Independence Model}

CRA quantifies the likelihood and impact of borrower default at both single-exposure and portfolio levels. The core ingredients are the probability of default (PD) and the loss given default (LGD), which combine to determine expected and tail losses over a horizon. Since debtors are not independent, portfolio models introduce latent risk factors to capture dependence (e.g., one-factor Vasicek or multi-factor structures, often paired with copulas for joint tails). The accurate simulation of correlated default scenarios and loss distributions is fundamental for practical tasks, e.g., pricing and capital allocation, as well as Basel stress testing. The principal computational bottleneck is the generation of high-fidelity samples from these correlated models and the evaluation of portfolio losses across many scenarios.

These requirements naturally favor latent-factor formulations that balance realism with tractability. The Gaussian conditional-independence (GCI) \emph{uncertainty model}~\cite{doi:10.1142/S021902491550034X} provides an analytically convenient link from shared risk factors to debtor-level default probabilities and is widely used as a classical baseline~\cite{Vasicek2002,Gordy2003}. Classically, GCI posits that the debtor-level conditional default probabilities are conditioned on one (or several) latent normal risk factor(s) $z$ as:
\begin{equation}
PD_k(z)
=\Phi\!\left(
\frac{\Phi^{-1}(p_k^{0})-\sqrt{\rho_k} z}{\sqrt{1-\rho_k}}
\right),
\label{eq:gci-pd}
\end{equation}
where $\Phi$ is the standard normal Cumulative Distribution Function (CDF), $p_k^{0}$ and $\rho_k$ are, respectively, the baseline default level and the correlation of the k-th debtor (asset). 

For the quantum mapping of this uncertainty model, we define the conditional default probability
$PD_k(z)$ as the probability
of measuring the state $\ket{1}$ on a target qubit,
\begin{equation}
    PD_k(z) \equiv P_1 = \sin^2\!\big(\alpha_k z + \beta_k\big).
    \label{eq:pd_sin2}
\end{equation}
Here, the parameters $\alpha_k$ and $\beta_k$ are obtained by
representing the classical conditional default probability function in a linear form within the latent
factor $z$, such that the corresponding quantum rotation angle depends affinely
on $z$ with a slope $\alpha_k$ and an offset $\beta_k$. Such parameters have been computed analytically, and a detailed explanation on the calculation is available in Supplementary Material (Formula 5 for $\beta_k$ and 6 for $\alpha_k$ of Supplementary Material). In this way,
conditional default mechanisms can be implemented by parameterizing single-qubit rotations of the type
\begin{equation}
    R_y\big(2(\alpha_k z + \beta_k)\big),
\end{equation}
so that the amplitude of $\ket{1}$ encodes the desired conditional default
probability.

In the circuit, the latent factor $z$ is represented by an $n$-qubit register,
whose computational basis states encode a discrete set of latent
factor values.
Absorbing these discrete values into effective linear parameters with slope
$\tilde{\alpha}_k$ and offset $\tilde{\beta}_k$, the gate implemented on the
target qubit can be written compactly as
\begin{equation}
    R_y\big(2(\tilde{\alpha}_k \,\hat{z}_{\text{code}} + \tilde{\beta}_k)\big),
    \label{eq:ry_zcode}
\end{equation}
where $\hat{z}_{\text{code}}$ denotes the integer value encoded by the
$n$-qubit $z$-register and controls the rotation angle, thereby encoding the
conditional default probabilities $PD_k(z)$ across the encoded configurations
(see Supplementary Material for further details).

This setting is typical of a single risk factor model, but in general the modeling requests the encoding of more of them. The total number of qubits required for encoding the GCI model is, in fact, determined by two contributions: the encoding of assets default event and the representation of the risk factors. In the present setting, each asset default event is encoded with one qubit, so the number of qubits related to this purpose scales linearly with the number of assets in the portfolio. In addition, each risk factor requires a dedicated register to load its corresponding normal distribution. The size of this register depends on the chosen discretization, which plays a crucial role in the accuracy of the model: allocating more qubits to a risk-factor register allows for a finer discretization, and therefore for a larger number of bins to represent the corresponding distribution. As a result, the total qubit cost depends not only on the number of assets and risk factors, but also on the level of discretization used for each risk factor, as:
\begin{equation}
N_{\mathrm{qubits}} = N_{\mathrm{assets}} + \sum_{k=1}^{N_{\mathrm{risk}}} n_k,
\end{equation}
where $N_{\mathrm{assets}}$ is the number of qubits used to encode the assets, $N_{\mathrm{risk}}$ is the number of risk factors, and $n_k$ is the number of qubits in the register used to discretize the $k$-th risk factor. In this work, we focused on a scenario involving one asset and one latent factor employing three qubits: $q_0$ and $q_1$ form the $z$-register ($n=2$) that prepares a discrete normal distribution over the latent factor $z$, while $q_2$ is the asset qubit that encodes the conditional default probability. As an example, scaling the model to two assets and two risk factors requires at least 6 qubits. Two qubits are used to represent the assets default event, while each risk factor is encoded with \(n_k = 2\) qubits, allowing every discretized normal distribution to be represented with \(2^{n_k} = 4\) bins. \\In real credit-risk applications, portfolio size can range from hundreds to many thousands of exposures, while the number of systematic risk factors depends on the modeling framework. This shows directly that moving from toy models to real business cases can quickly increase the hardware requirements. For example, a portfolio with $1000$ assets already requires $1000$ qubits for the assets' default encoding alone, and adding 10 risk factors, discretized with $6$ qubits each, increases the total to $1060$ qubits. \newline

\begin{figure}[h]
	\centering
    \includegraphics[width=0.8\columnwidth]{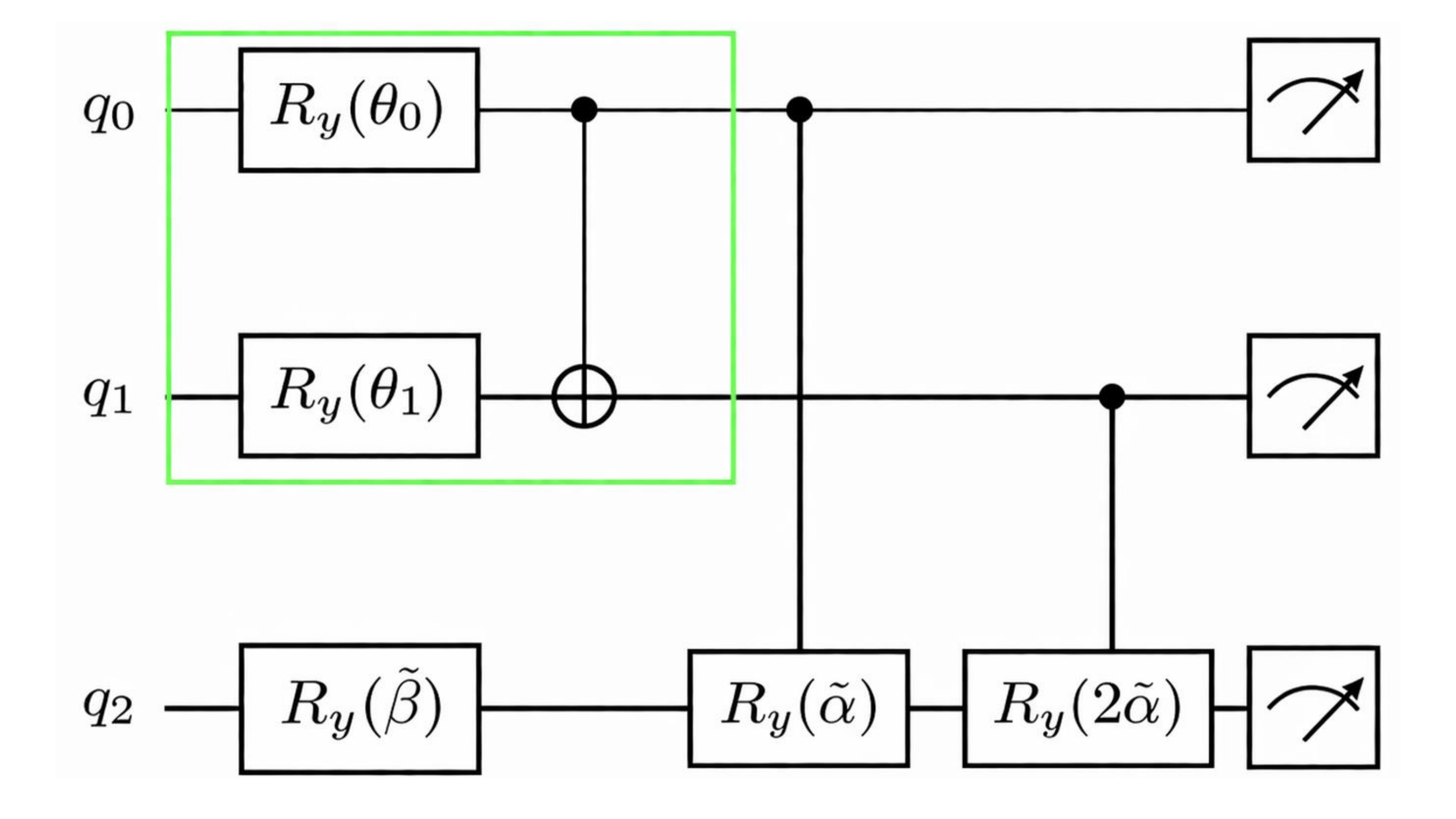}
    \caption{Quantum circuit for the target GCI model with one asset and one risk factor. 
    The green box underlines the Gaussian loading sub-circuit.}
    \label{fig:linrot-circuit}
\end{figure}

In the one-risk-factor/one-asset scenario, the loading of the latent factor uses two single-qubit \(R_y\) rotations followed by a \(\mathrm{CNOT}_{0,1}\) entangler, shaping the two-qubit amplitudes to approximate a Gaussian over the \((q_0,q_1)\) register (the green square in Fig.~\ref{fig:linrot-circuit}). The state of the asset (default) qubit \(q_2\) can be prepared via controlled \(R_y\) rotations conditioned on \(q_0\) and \(q_1\), such that the effective rotation angle, and consequently the probability of measuring \(\ket{1}\) on \(q_2\), depends on the encoded value of the normal factor. This realizes the conditional map
\(PD(z)\approx \sin^{2}(\alpha z+\beta)\), linking the risk factor to the asset’s default probability within a shallow, hardware-efficient circuit.

\subsection{Variational circuit for a normal distribution}\label{subsec:VCN}

A crucial role in the CRA and GCI models is played by the preparation of a discrete normal distribution over the latent factor. In fact, the GCI quantum circuit in Fig.~\ref{fig:linrot-circuit} includes a Gaussian loading sub-circuit on a register of $2$ qubits (green box). The generalization to $n$ qubits, spanning \(2^{n}\) computational basis states, yields the requirement to prepare a quantum state $\psi(\boldsymbol{\theta})$. According to~\cite{loaddistrubition},
\begin{equation}
  \lvert \psi(\boldsymbol{\theta}) \rangle \;=\; \sum_{b=0}^{2^{n}-1} \sqrt{p_b(\boldsymbol{\theta})}\,\lvert b\rangle,
\end{equation}
so that measuring in the computational basis returns the bitstring \(b\) with a probability of \(p_b(\boldsymbol{\theta})\).

We first adopt a distribution-agnostic strategy: select a hardware-efficient ansatz and train its parameters \(\boldsymbol{\theta}\) so that the model distribution \(p(\boldsymbol{\theta})\) matches a prescribed histogram
\begin{equation}
  p^{\star}=\{p_b^{\star}\}_{b=0}^{2^n-1}.
\end{equation}
Given a discrepancy functional \(D\!\big(p(\boldsymbol{\theta})\,\|\,p^{\star}\big)\), gradients are estimated via the parameter–shift rule, and \(\boldsymbol{\theta}\) is updated with a stochastic optimizer such as Adam~\cite{qgan,Kingma2014AdamAM}, \cite{Ballarin2025}, which is preferable compared to other optimizers for variational quantum algorithms~\cite{Pellow2021}, since it uses more direct and stable gradient-based updates, leading to faster
and smoother convergence when the optimization landscape is simple.

This recipe—compact ansatz, target histogram, shift-rule gradients, and Adam updates—supports arbitrary discrete distributions over the register.

As concrete instances, we synthesize discrete Gaussians on two and three qubits.

\begin{figure}[h]
\centering
\includegraphics[width=0.8\columnwidth]{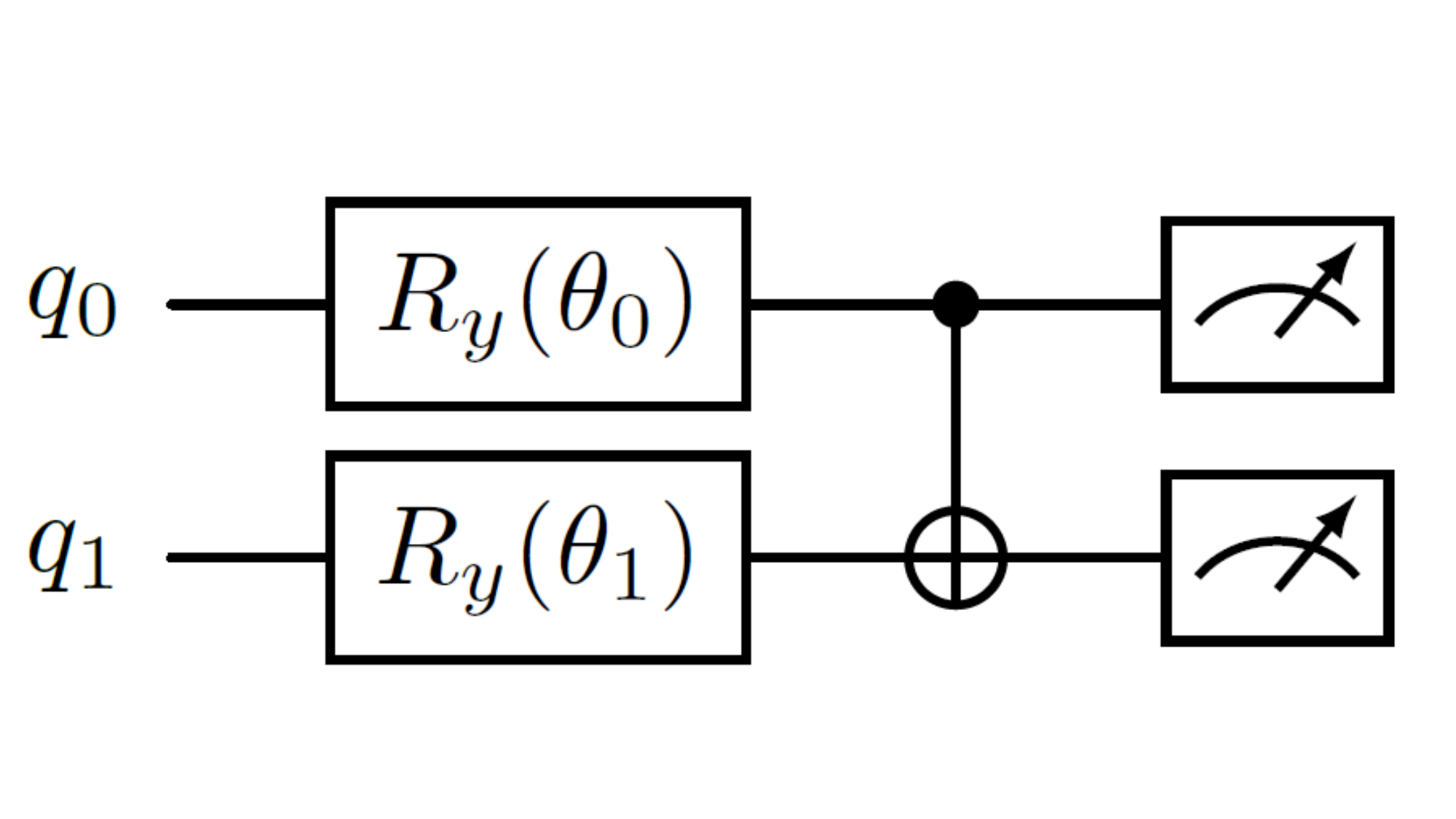}
\caption{Two-qubit quantum circuit for the Gaussian distribution loading.}
\label{TwoGauss}
\end{figure}

\textbf{Two qubits.}
On the register \((q_0,q_1)\), we use the hardware-efficient ansatz: single-qubit \(R_y\) rotations on each qubit followed by one entangler \(\mathrm{CNOT}_{0,1}\) (Fig.~\ref{TwoGauss}).
Let \(\boldsymbol{\theta}\) collect all rotation angles; the Born probabilities are
\begin{equation}
  p_b(\boldsymbol{\theta})=\bigl|\langle b\mid\psi(\boldsymbol{\theta})\rangle\bigr|^2,\qquad b\in\{0,1,2,3\}.
\end{equation}

\begin{figure}[h]
\centering
\includegraphics[width=0.8\columnwidth]{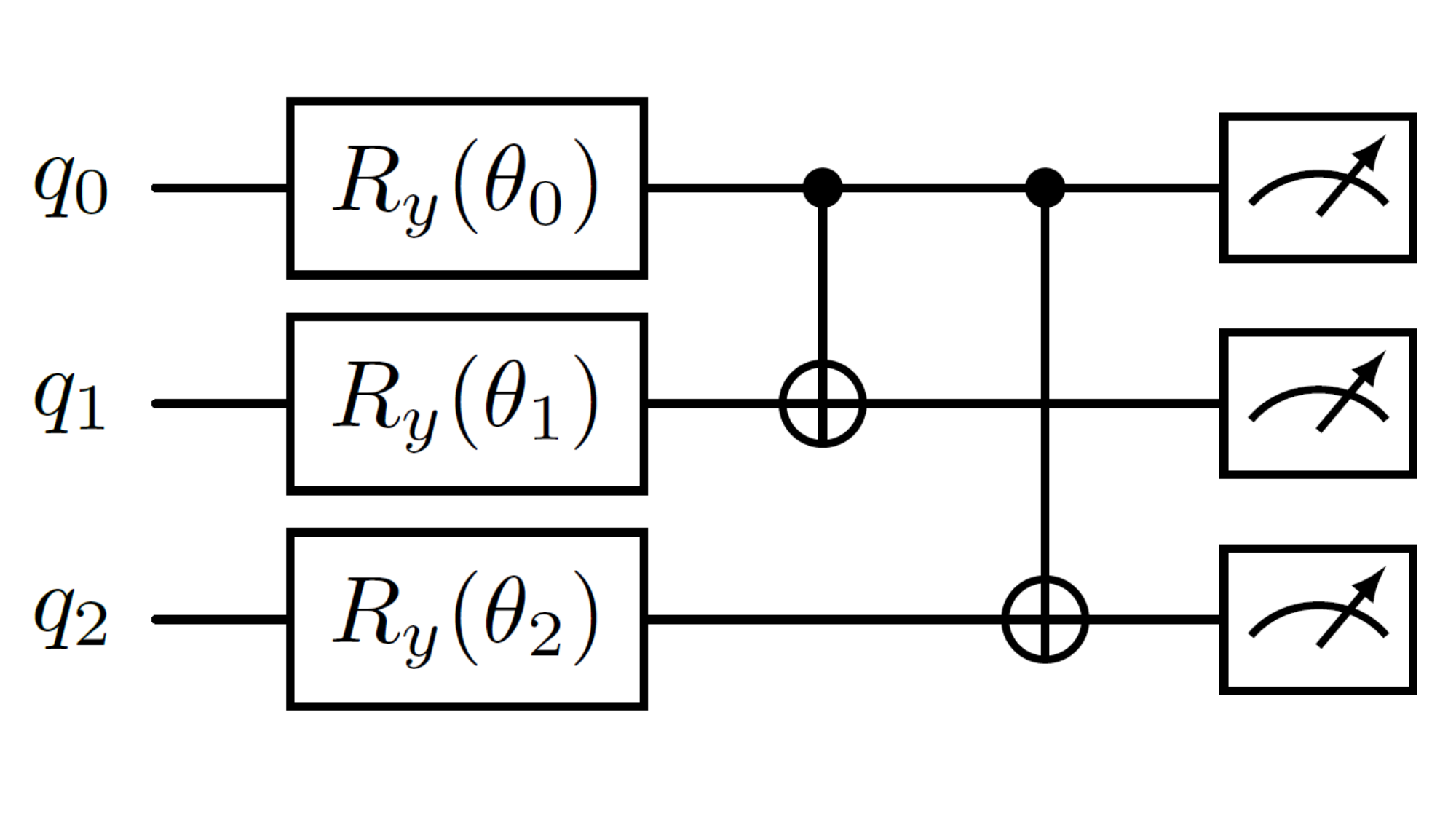}
\caption{Three-qubit quantum circuit for the Gaussian distribution loading.}
\label{ThreeGauss}
\end{figure}

\textbf{Three qubits.}
We extend the circuit to \((q_0,q_1,q_2)\) by adding an \(R_y(\theta_2)\) on \(q_2\) and a second entangler \(\mathrm{CNOT}_{0,2}\) (control \(q_0\)) (Fig.~\ref{ThreeGauss}).
The output probabilities are
\begin{equation}
  p_b(\boldsymbol{\theta})=\bigl|\langle b\mid\psi(\boldsymbol{\theta})\rangle\bigr|^2,\qquad b\in\{0,1,\dots,7\}.
\end{equation}
\textbf{Target and training.}
For either \(n\in\{2,3\}\) qubits, we map each basis index \(b\) to a grid point
\(z(b)\in[-z_{\max},z_{\max}]\) via an affine transform and define the target histogram of
\(\mathcal{N}(\mu,\sigma^2)\) as
\begin{equation}
  p_b^{\star}\propto \exp\!\Big(-\frac{(z(b)-\mu)^2}{2\sigma^2}\Big),\qquad b=0,\ldots,2^{n}-1.
\end{equation}
Parameters are learned by minimizing the distributional loss
\begin{equation}
  \mathcal{L}(\boldsymbol{\theta})=\sum_{b=0}^{2^{n}-1}\Big(p_b(\boldsymbol{\theta})-p_b^{\star}\Big)^2,
\end{equation}
using parameter-shift gradients and Adam updates.

After convergence, the measured histograms for the two- and three-qubit circuits closely match the target normal specified by \((\mu,\sigma)\), yielding shallow, hardware-efficient loaders suitable for subsequent calibration and hyperparameter tuning on the sQPU.

\subsection{Hardware-Level normal distribution preparation: two and three Qubits}

Starting with a two-qubit system, we study how the gates affect the probability amplitudes of the basis 
\(\{|00\rangle, |01\rangle, |10\rangle, |11\rangle\}\).
The single–qubit rotation
\begin{equation}
R_{Y}(\theta)=
\begin{bmatrix}
\cos(\theta/2) & -\sin(\theta/2)\\
\sin(\theta/2) & \phantom{-}\cos(\theta/2)
\end{bmatrix}
\end{equation}
acts on each qubit. From the initial state \(|00\rangle\), whose amplitude vector is \([1,0,0,0]^{\top}\), we obtain
\begin{equation*}
\begin{array}{c}
|00\rangle \\ |01\rangle \\ |10\rangle \\ |11\rangle
\end{array}
\;\;\;
\left[\begin{array}{c} 1 \\ 0 \\ 0 \\ 0 \end{array}\right]
\;\xrightarrow{\,R_Y(\theta_{0})\otimes R_Y(\theta_{1})\,}\;
\left[\begin{array}{c}
\cos(\theta_{0}/2)\cos(\theta_{1}/2)\\
\cos(\theta_{0}/2)\sin(\theta_{1}/2)\\
\sin(\theta_{0}/2)\cos(\theta_{1}/2)\\
\sin(\theta_{0}/2)\sin(\theta_{1}/2)
\end{array}\right].
\end{equation*}
Applying a CNOT with control \(q_0\) and target \(q_1\) yields
\begin{equation}
\left[\begin{array}{c}
\cos(\theta_{0}/2)\cos(\theta_{1}/2)\\
\cos(\theta_{0}/2)\sin(\theta_{1}/2)\\
\sin(\theta_{0}/2)\sin(\theta_{1}/2)\\
\sin(\theta_{0}/2)\cos(\theta_{1}/2)
\end{array}\right],
\label{eq:two_qubit_after_cnot}
\end{equation}
i.e., the amplitudes of \(|10\rangle\) and \(|11\rangle\) are swapped. Consequently, the output probabilities depend on the choices of \(\theta_{0}\) and \(\theta_{1}\).

For a symmetric, bell-shaped target (e.g., a discrete Gaussian), we require
\[
P_{|00\rangle}=P_{|11\rangle},\qquad
P_{|01\rangle}=P_{|10\rangle},\qquad
P_{|00\rangle,|11\rangle} < P_{|01\rangle,|10\rangle}.
\]
From Eq.~\eqref{eq:two_qubit_after_cnot}, these symmetries are satisfied when
\begin{equation}
\theta_{0}=\pm\frac{(2n_{0}+1)\pi}{2},
\theta_{1}=2\pi n_{1}\pm\frac{(2n_{0}+1)\pi}{2},
 n_{0},n_{1}\in\mathbb{N},
\end{equation}
and the “central-mass” condition \(P_{|00\rangle,|11\rangle}<P_{|01\rangle,|10\rangle}\) further imposes
\begin{equation}
\cos(\theta_{1}/2)<\sin(\theta_{1}/2)\;\;\Longleftrightarrow\;\; \pi/2<\theta_{1}<3\pi/2\;\;(\text{mod }2\pi).
\end{equation}
By taking the software-optimized loader as our baseline, we can then fix \(\theta_{0}\) as the reference angle and tune \(\theta_{1}\) to match the characteristics of the quantum hardware at hand. The output profile is uniform when $\theta_{1}=\{\pi/2,\,5\pi/2,\ldots\}$, while to obtain a Gaussian-like shape, one can choose \(\theta_{1}\) so that \(\cos(\theta_{1}/2)<\sin(\theta_{1}/2)\), which yields \(P_{|00\rangle,|11\rangle}<P_{|01\rangle,|10\rangle}\).

\textbf{Extension to Three Qubits.}
We now scale to a three-qubit register with basis
\(\{|000\rangle, |001\rangle, |010\rangle, |011\rangle, |100\rangle, |101\rangle, \\|110\rangle, |111\rangle\}\).
Starting from the two-qubit output (with \(\theta_{0}=\pi/2\)) and adding an \(R_Y(\theta_{2})\) on \(q_2\), then a second entangler \(\mathrm{CNOT}_{0,2}\) (control \(q_0\), target \(q_2\)), the amplitude trajectory reads
\begin{equation*}
\frac{1}{\sqrt{2}}
\begin{bmatrix}
\cos(\theta_{1}/2)\\ 0\\ \sin(\theta_{1}/2)\\ 0\\
\sin(\theta_{1}/2)\\ 0\\ \cos(\theta_{1}/2)\\ 0
\end{bmatrix}
\;\xrightarrow{\,R_Y(\theta_{2})\,}\;
\frac{1}{\sqrt{2}}
\begin{bmatrix}
\cos(\theta_{1}/2)\cos(\theta_{2}/2)\\
\cos(\theta_{1}/2)\sin(\theta_{2}/2)\\
\sin(\theta_{1}/2)\cos(\theta_{2}/2)\\
\sin(\theta_{1}/2)\sin(\theta_{2}/2)\\
\sin(\theta_{1}/2)\cos(\theta_{2}/2)\\
\sin(\theta_{1}/2)\sin(\theta_{2}/2)\\
\cos(\theta_{1}/2)\cos(\theta_{2}/2)\\
\cos(\theta_{1}/2)\sin(\theta_{2}/2)
\end{bmatrix}
\end{equation*}

\begin{equation*}
\xrightarrow{\,\mathrm{CNOT}_{0,2}\,}\;
\frac{1}{\sqrt{2}}
\begin{bmatrix}
\cos(\theta_{1}/2)\cos(\theta_{2}/2)\\
\cos(\theta_{1}/2)\sin(\theta_{2}/2)\\
\sin(\theta_{1}/2)\cos(\theta_{2}/2)\\
\sin(\theta_{1}/2)\sin(\theta_{2}/2)\\
\sin(\theta_{1}/2)\sin(\theta_{2}/2)\\
\sin(\theta_{1}/2)\cos(\theta_{2}/2)\\
\cos(\theta_{1}/2)\sin(\theta_{2}/2)\\
\cos(\theta_{1}/2)\cos(\theta_{2}/2)
\end{bmatrix}. 
\end{equation*}
The combined action of \(\mathrm{CNOT}_{0,1}\) and \(\mathrm{CNOT}_{0,2}\) (control \(q_0\) in both cases) swaps the pairs
\(|100\rangle \leftrightarrow |101\rangle\) and \(|110\rangle \leftrightarrow |111\rangle\).
Under the same symmetry conditions on \(\theta_{0}\) and \(\theta_{1}\) as above, a symmetric three-qubit profile requires
\[
P_{|000\rangle}=P_{|111\rangle},\;\;
P_{|001\rangle}=P_{|110\rangle},\;\;
P_{|010\rangle}=P_{|101\rangle},\;\;\]
\\
\[P_{|011\rangle}=P_{|100\rangle},
\]
which is obtained when
\begin{equation}
\theta_{2}=2\pi n_{2}\pm\theta_{1},\qquad n_{2}\in\mathbb{N}.
\end{equation}
To enforce a central tendency (highest mass near \(|000\rangle\) and \(|111\rangle\)) with strictly decreasing “rings”,
\[
P_{|000\rangle,|111\rangle} \;<\; P_{|001\rangle,|110\rangle} \;<\;
P_{|010\rangle,|101\rangle} \;<\; P_{|011\rangle,|100\rangle},
\]
one further selects \(\theta_{2}\) so that \(\cos(\theta_{2}/2)<\sin(\theta_{2}/2)\), i.e.
\begin{equation}
\pi/2<\theta_{2}<5\pi/2\quad(\text{mod }4\pi).
\end{equation}
These analytic angle relations provide hardware-level guidance for shaping symmetric, bell-like histograms with shallow entangling patterns on today’s sQPUs. In fact, rotation angles in sQPUs can be controlled by adjusting the amplitude and the shape of microwave pulses at room temperature with a dedicated FPGA-based electronics~\cite{Riste2020}. As we will show in Sec.~\ref{sec:results}, a variational optimization of such electrical parameters allows for a hardware-aware hypertuning of rotation angles, first optimized classically through quantum circuit simulations. This method eventually counteracts for systematic errors arising at the electronics level, such as amplitude and phase-errors, and automatically takes into account the intrinsic noisy nature of current NISQ processors, as well as the electrodynamics of different qubits in the same processor. Therefore, in Sec.~\ref{sec:results}, we examine how sQPU connectivity and circuit choices (e.g., entangler placement and single-qubit rotation settings) affect the quality of the prepared distribution.

\section{Hardware description}
\label{sec:hardware}

\begin{figure}[h!]
\centering
\includegraphics[width=0.5\textwidth]{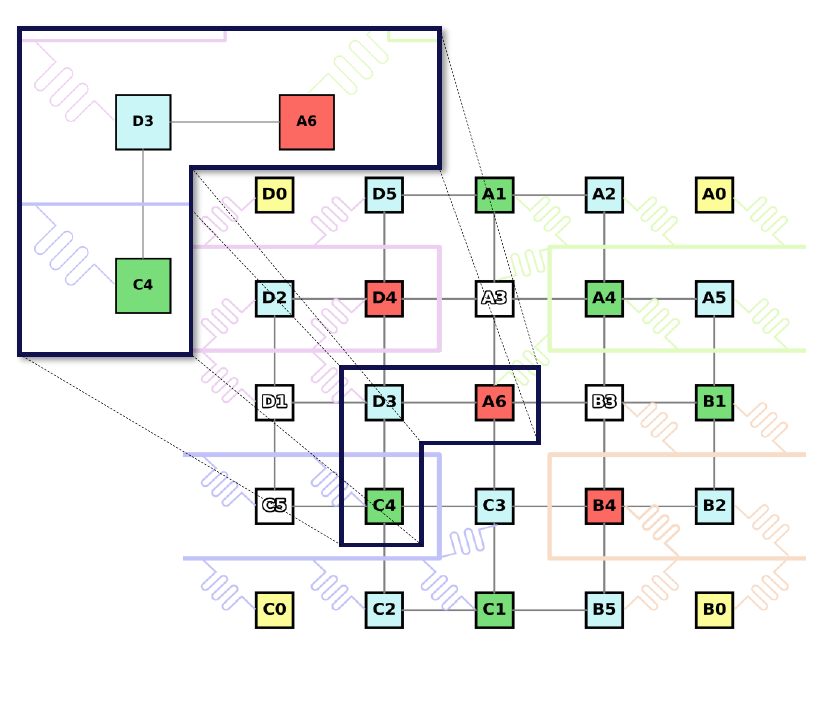}
\caption{Schematics of the processor, with focus on three qubit register A6, D3, C4. The processor has 4 feedlines for multiplexed readout: feedline A (light green), coupled to $6$ qubits, and feedlines B (pink), C (indigo), and D (purple), each coupled to $5$ qubits. In green: low frequency qubits; in blue: mid frequency qubits; in red: high frequency qubits; in yellow: isolated qubits. White color identifies qubits that are not operational.}
\label{processor}
\end{figure}

The superconducting Quantum Processing Unit (sQPU) used in this work is a Contralto-D from Quantware (Fig.~\ref{processor}), composed of $17$ operational symmetric flux-tunable transmons, connected in a rectangular 2D matrix through fixed high-frequency bus resonators.
\begin{center}
	\begin{table*}[th!]
		\centering
		\begin{tabular}{|c|c|c|c|c|c|c|}
			\hline
			\rule{0pt}{3ex}     & {\bfseries $T_{1}\,(\mu s)$} & {\bfseries $T^{*}_{2}\,  (\mu s)$} & {\bfseries $T_{2E}\, (\mu s)$} & $F_{avg}\,(\%)$ & $F_{RO}\,(\%)$ & $F_{CZ}\,(\%)$ \\
			\hline
			\rule{0pt}{3ex} \textcolor{red}{{\bfseries A6}} &  43 $\pm$ 1 & 37 $\pm$ 3 & 40 $\pm$ 1  & 99.855 $\pm$ 0.005 & 0.95 $\pm$ 0.01 & \\
			\hline
			\rule{0pt}{3ex} \textcolor{green}{{\bfseries C4}} & 38 $\pm$ 1 & 20 $\pm$ 1  & 42 $\pm$ 4 & 99.924 $\pm$ 0.002 & 0.95 $\pm$ 0.01 &\\
			\hline
			\rule{0pt}{3ex} \textcolor{blue}{{\bfseries D3}} & 32 $\pm$ 1 & 20 $\pm$ 6 & 37 $\pm$ 6 & 99.91 $\pm$ 0.01 & 0.94 $\pm$ 0.01  & \\
            \hline
            \rule{0pt}{3ex} \textcolor{red}{{\bfseries A6}}-\textcolor{blue}{{\bfseries D3}} &  &  &  &  &   & 99.7 $\pm$ 0.2 \\
            \hline
            \rule{0pt}{3ex} \textcolor{blue}{{\bfseries D3}}-\textcolor{green}{{\bfseries C4}} &  &  &  &  &   & 99.05 $\pm$ 0.07\\
			\hline			
		\end{tabular}
		\caption{Contralto-D QPU specifications for A6, D3, C4 qubits: coherence times ($T_1$ - relaxation time, $T_2^{*}$ - Ramsey coherence time, $T_{2E}$ - Hahn-Echo coherence time, with errors provided by $100$ repeated experiments), average single-qubit gate ($F_{avg}$) and readout ($F_{RO}$) fidelities, and two-qubit CZ interleaved average gate fidelity $F_{CZ}$ for the pairs. In green low frequency band qubits, in blue mid frequency band qubits and in red high frequency band qubits. }
		\label{tabbco}
	\end{table*} 
\end{center} 
\begin{table*}[th!]
	\centering
	\includegraphics[width=\textwidth]{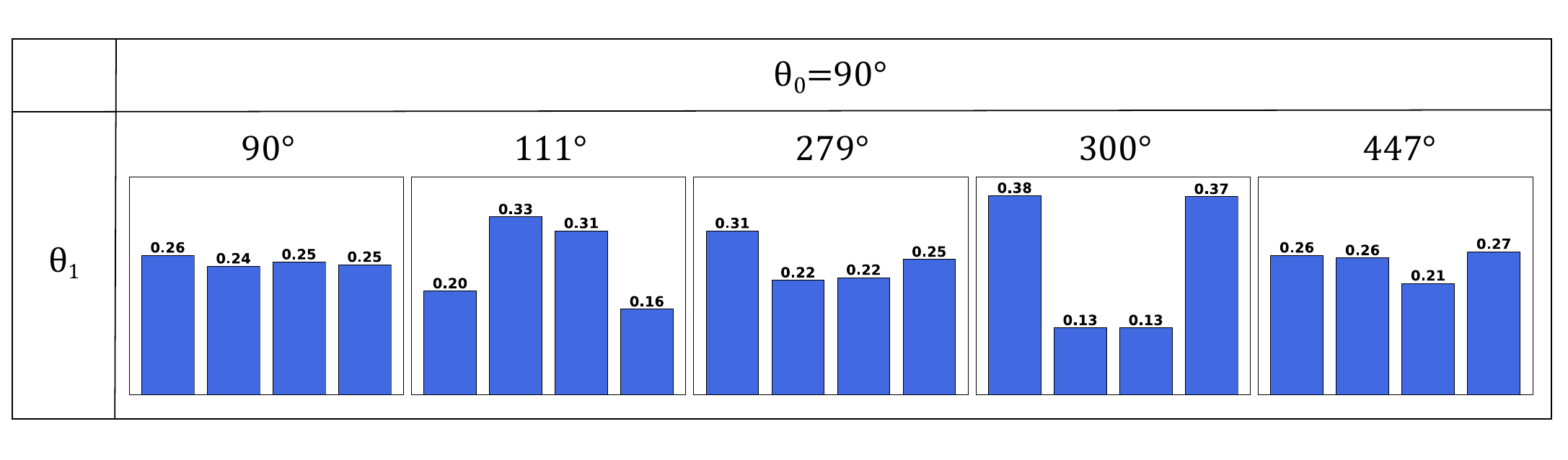}
	\caption{Two-qubit circuit output on D3-C4 register as a function of $\theta_{1}$ and fixed $\theta_{0}=90^{\circ}$. For $\theta_{1} \in \left\{90^{\circ},270^{\circ}\right\}$ probability amplitudes feature a gaussian-like shape, while for $\theta_{1} \in \left\{270^{\circ},450^{\circ}\right\}$ the concavity is inverted.}
	\label{GeneralTrend}
\end{table*}
Qubits are designed to fall into three frequency bandwidths: high-frequency qubits have frequencies of the order of $6\;GHz$ (in red), medium frequency qubits have frequencies of the order of $5\;GHz$ (in cyan), and low frequency qubits lie in a frequency range of the order of $4\;GHz$ (in green). All the qubits are equipped with a dedicated superconducting readout resonator, equally distributed over $4$ readout feedlines, thereby guaranteeing multiplexed readout. They also use dedicated drive and flux lines to implement single-qubit X-Y gates and Z-gates, as well as the native two-qubit gate of the sQPU, the Conditional-Z (CZ) gate, which we implement by using Sudden-Net-Zero (SNZ) flux pulses on the highest-frequency qubit in a pair~\cite{Negirnac2021}. Flux lines also allow for parking the qubits at specific working points by using static magnetic flux. In this work, we have focused on qubits C4, D3, and A6 (low-, medium-, and high-frequency qubits, respectively), which have been operated at their flux sweet-spots. The remaining qubits of the matrix have been operated at their anti-sweet spot, i.\,e., nominally at zero frequency, to limit the effect of frequency crowding and qubit crosstalk.

The sQPU is installed at the coldest stage of a Bluefors XLD1000SL, equipped with cryogenic attenuated drive and read-in lines, superconducting cryogenic coaxial flux lines, and read-out lines with High-Electron Mobility Transistors (HEMT) amplifiers, anchored at the 4K plate of the cryostat, and a double junction isolator at the MXC. Additional detail on the cryogenic experimental setup is reported in the Supplementary Material of Ref.~\cite{Ahmad2025_magic}. At room temperature, the sQPU interfaces with a host PC through LAN connections via a Qblox cluster, equipped with FPGA-based modules for control, read-out, and flux pulsing of qubits. Static flux-pulses for qubits parking are delivered by low-noise DC electronics, the SPI rack, and combined with flux-pulsed signals from the cluster using bias-tees at room temperature. The Qblox electronics can be controlled by a Python-based framework, \emph{Quantify}~\cite{Quantify}, which allows us to design experimental schedules to run quantum circuits of varying complexity, control every physical quantity of X, Y, Z, and decompose the required two-qubit CNOT gate in terms of the native CZ gate directly at the pulse-level.  Specifically, in this sQPU, X and Y gates are implemented using Derivative Reduction Adiabatic Gate (DRAG) microwave pulses of $60\;ns$: rotation angles are set by changing the amplitude of such pulses, by using as an input the rotation angle in degree directly dependent on the amplitude value, while the rotation axis is selected by changing the phase of the microwave signals (again, setting an angle in degree, directly connected to the amount of phase-delay in the microwave signal). Microwave pulses are also used to excite the dedicated superconducting resonators, and readout is performed by digitizing the output signal through the readout modules. Meanwhile, Z single-qubit gates are virtually implemented by phase updates of the control signal~\cite{Zvirtual,McKay2017}. All other single-qubit rotations are generally decomposed in terms of the X, Y, and Z rotations. Additional details can be found in~\cite{Krantz2019}. 

In Tab.~\ref{tabbco}, we report on the coherence times, gate, and read-out fidelities of the investigated qubits, acquired right before the experiments to have available benchmarking coherence and fidelity parameters of the NISQ machine. This comes in handy when we want to investigate any relation between a quantum algorithm output and the noise of the NISQ processor~\cite{Ahmad2025_magic}. Additional details on calibration procedures and the gate and readout fidelities, as well as the assignment calibration matrices for the experiments reported in this work, are collected in the Supplementary Material.

\section{Results}
\label{sec:results}

\subsection{Two-qubit normal distribution encoding and loading}
\label{sec:results-norm_dist}

\begin{figure}[t!]
	\centering
\includegraphics[width=0.9\columnwidth]{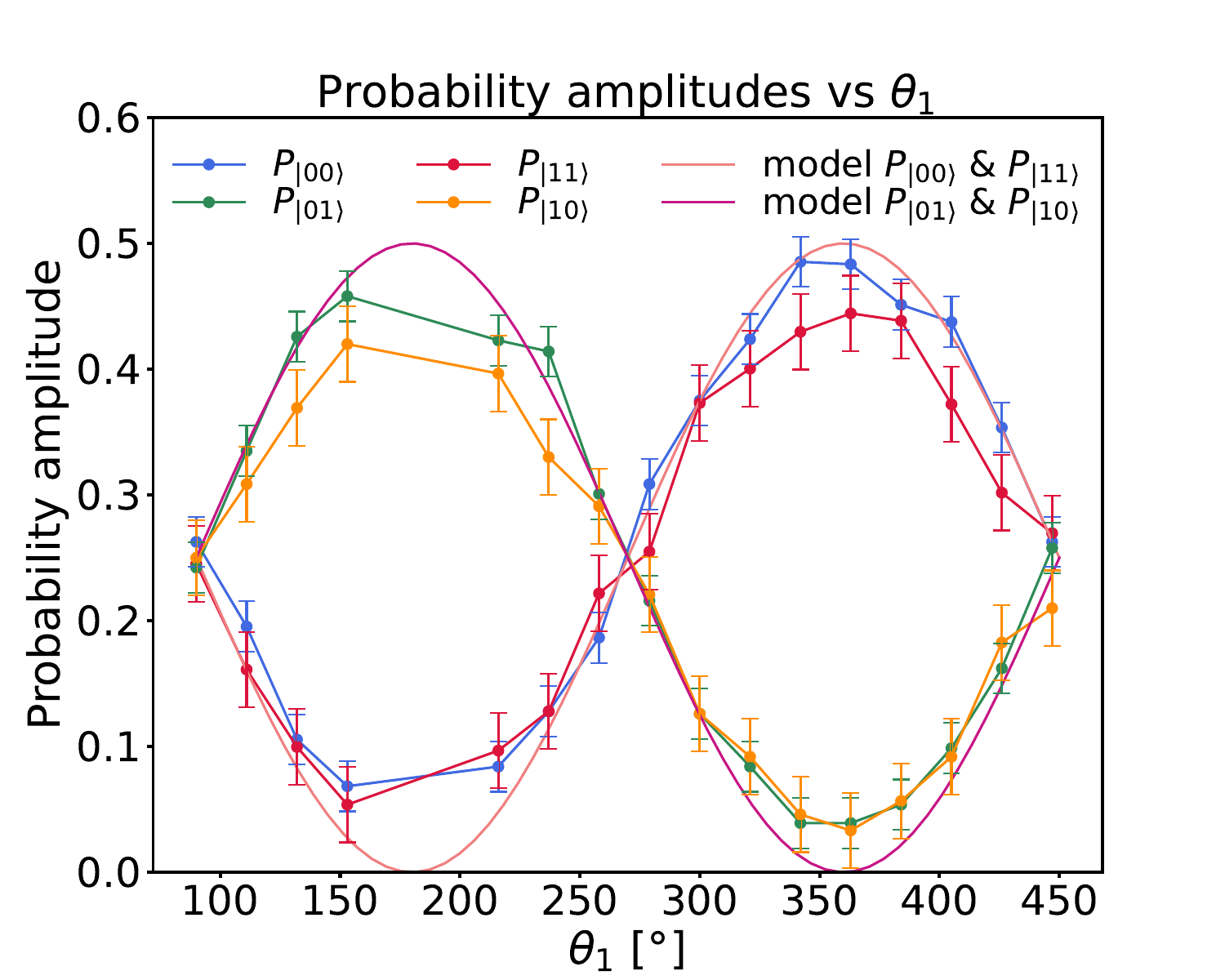}
	\caption {Comparison between probability amplitudes of $|00\rangle,|01\rangle,|10\rangle,|11\rangle$ states and theoretical model as a function of $\theta_{1} \in \left\{90^{\circ},450^{\circ}\right\}$ and fixed $\theta_{0}=90^{\circ}$. }
	\label{ProbvsTheta}
\end{figure}

The first experimental investigation involved the D3-C4 pair, and the goal was to identify the best $\theta_{0}, \theta_{1}$ angles combination to achieve a two-qubit standard normal distribution. By preparing the two-qubit system in the $|00\rangle$ state, we first applied an \(R_y\) gate on D3 with fixed $\theta_{0}=90^{\circ}$ and $\theta_{1}=111^{\circ}$, obtained through the target and training procedure and the Adam optimization discussed in Sec.~\ref{subsec:VCN}. The experimental result on the quantum machine is reported in Tab.~\ref{GeneralTrend}. With the goal of experimentally assessing at the hardware level the effect of the rotation angle on the probability distribution shape, we swept the angle $\theta_{1}$ for the \(R_y\) gate on C4 in the range $\left[90^{\circ},450^{\circ}\right]$ with a step of $21^{\circ}$. To provide a comparison with the initial experiment, in Tab.~\ref{GeneralTrend} we collect some of the acquired output distributions to show that they exhibit a periodic trend: for $90^{\circ}<\theta_{1}<270^{\circ}$, the probability amplitude distributions show a positive concave shape; for $270^{\circ}<\theta_{1}<450^{\circ}$, the concavity inverts, while for $\theta_{1}$ around $90^{\circ}$, $270^{\circ}$, and $450^{\circ}$, the distribution becomes uniform. The experimental probability amplitudes for the two-qubit basis states as a function of all the angles $\theta_1$ are also reported in Fig.~\ref{ProbvsTheta}, showing a reasonable agreement with the theoretical expectation.
	\begin{figure}[t!]
	\centering
		\includegraphics[width=0.9\columnwidth]{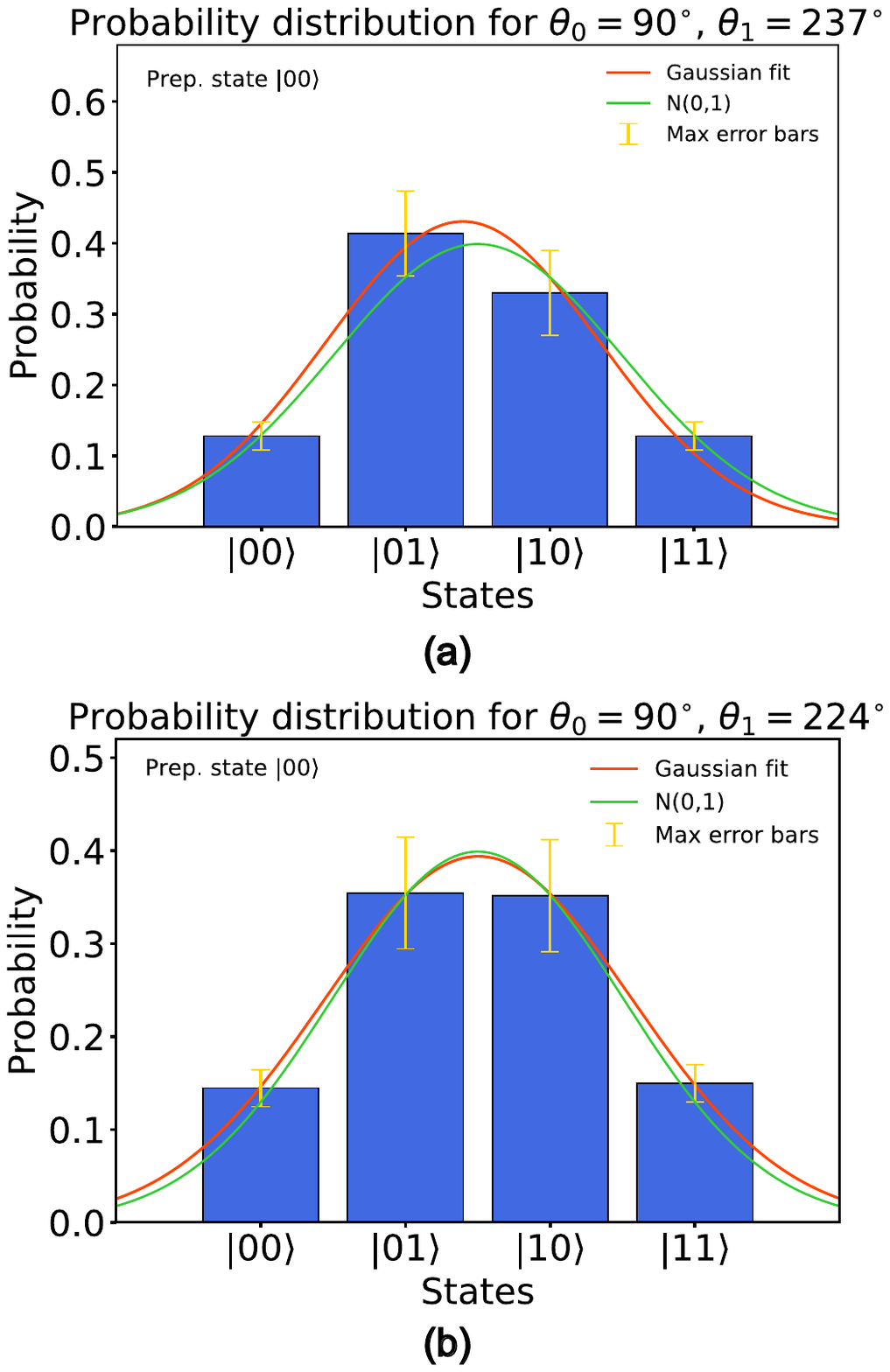} 
    \caption{Two-qubit Gaussian distribution at the optimal $\theta_0$ and $\theta_1$ angle and corresponding Gaussian fit for two different pairs. In (a), experimental results for pair D3-C4. In (b), experimental results for pair D3-A6. In green, the standard normal distribution is plotted for comparison.}
    \label{single_qubit_distribution}
\end{figure}
This first screening allowed us to identify a promising region with a positive concave shape and probabilities compatible with the simulated output. 
We repeated the experiment by thickening the step to $1^{\circ}$ between these angles to find the optimal $\theta_1$, which resulted in $\theta_{1}=237^{\circ}$ (Fig.~\ref{single_qubit_distribution} (a)). Similar experiments were also performed by preparing $|11\rangle$ as the initial state, which can be found in the Supplementary Material. 

As one can notice, the quantum circuit output can reveal fluctuations that may make the output distribution slightly asymmetric. This can be accounted for State Preparation And Measurement (SPAM) errors in an NISQ sQPU. Therefore, we have investigated such an error by performing a statistical analysis of $100$ measurements for the optimal set of angles. This provided us with a way to estimate the error bars on the quantum circuit output in terms of the population states. In detail, we assumed the deviation from the symmetric case, i.\,e.,  $P_{|00\rangle}=P_{|11\rangle}$ and $P_{|01\rangle}=P_{|10\rangle}$, so $\Delta P_{|00\rangle,|11\rangle}=P_{|00\rangle}-P_{|11\rangle}$ and $\Delta P_{|01\rangle,|10\rangle}=P_{|01\rangle}-P_{|10\rangle}$. The detailed analysis for one example output distribution can be found in the Supplementary Material. 

Finally, to investigate the role of hardware connectivity in this quantum algorithm, we performed the same experiments on the pair D3 and A6, with A6 (the highest-frequency qubit in the pair) as the target qubit. As discussed in the Supplementary Material, to optimize the performance of the CNOT gate between D3 and A6 when A6 is both the flux-tuned qubit and the target one in the register, a counter-phase Z-gate has been included in the CNOT decomposition to address depolarizing errors due to single-qubit phase-accumulation. The Gaussian distribution, obtained for an optimal angle $\theta_{1}=224^{\circ}$ (Fig.~\ref{single_qubit_distribution} (b)), differs from that obtained for D3-C4, and is more consistent with a standard normal distribution (highlighted in green). This result suggests that not only is it fundamental to optimize rotation angles on the quantum machine after prior optimization angles at the classical level, but also that the optimal rotation angles depend on the physical and hardware parameters of the particular pair of qubits considered in the quantum register. This dependence cannot be fully captured by offline hardware-agnostic optimization.
\begin{figure}[t]
	\centering
	\includegraphics[width=\columnwidth]{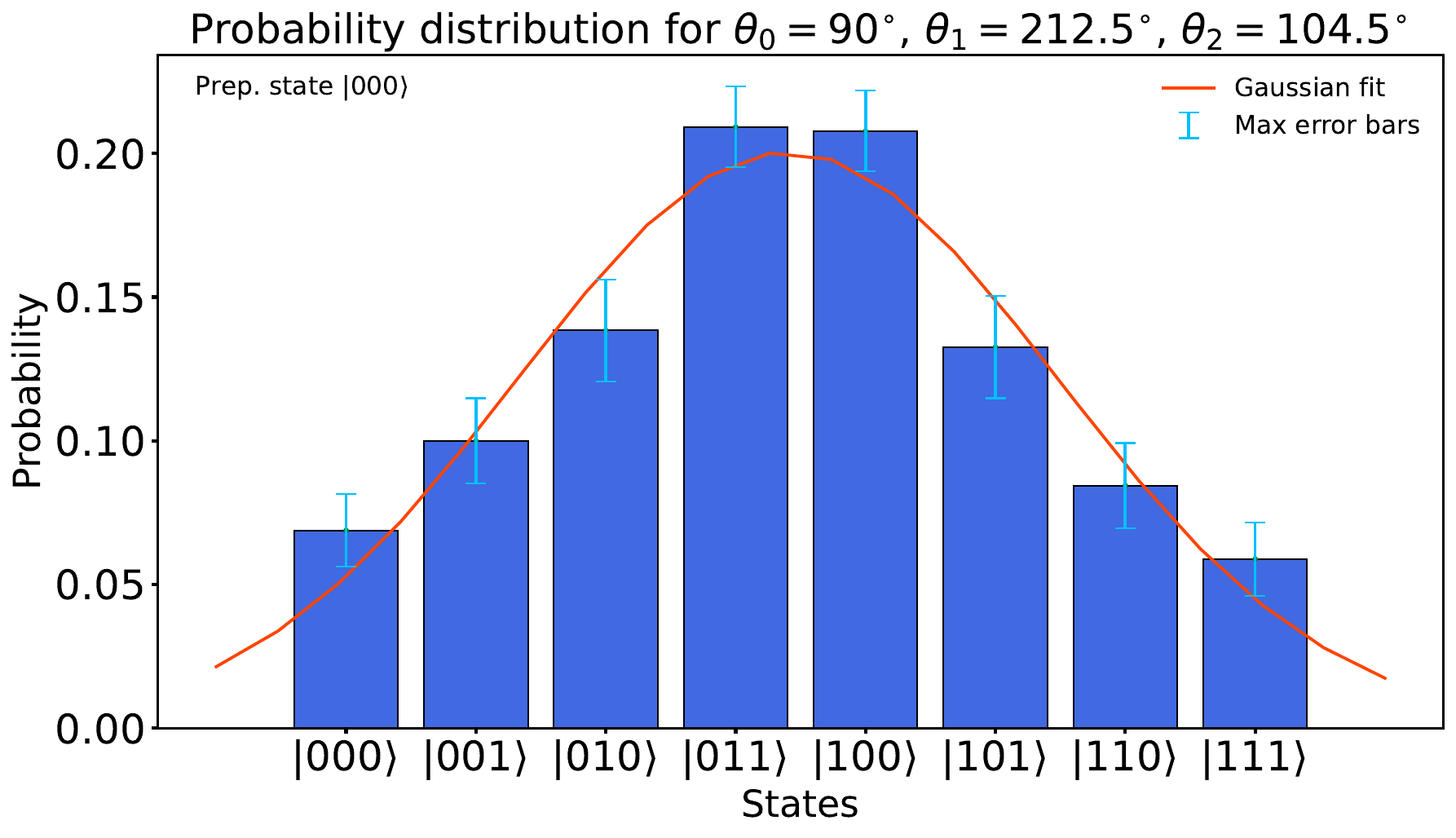}
	\caption{Three-qubit Gaussian distribution for qubits D3-A6-C4.}
	\label{3q}
\end{figure}
\begin{figure*}[t!]
	\centering
    \includegraphics[width=1\textwidth]{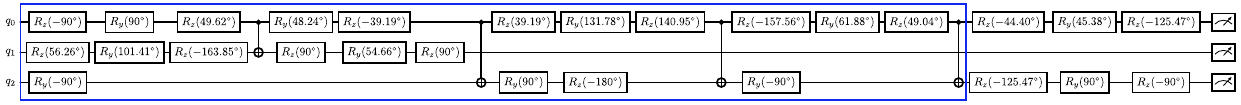}
	\caption{Use-case GCI transpiled quantum circuit. In the blue square, the sub-circuit provides an output equivalent to a three-qubit uniform distribution circuit.}
	\label{TC}
\end{figure*}

\subsection{Three-qubit normal distribution encoding and loading}
\label{sec:results-3qnorm_dist}

For the three-qubit distribution sampling, we used the triplet D3-A6-C4, where the implemented quantum circuit is reported in Fig.~\ref{ThreeGauss}. First, we fixed $\theta_{0}=90^{\circ}$ of the \(R_y\) gate on D3 and changed $\theta_{1}$ in the range $\left[90^{\circ}, 450^{\circ}\right]$ with a step of $36^{\circ}$ for the \(R_y\) gate on A6, and $\theta_{2}$ in the range $\left[90^{\circ}, 450^{\circ}\right]$ for the \(R_y\) gate on C4. Again, the goal was to study the output distributions as a function of the angle, also obtaining a transition from positive to negative concavity, passing through a uniform output distribution (see the Supplementary Material). Then, we adjusted the step on the angles in a range for which the output distribution better satisfies the requirements of a gaussian-like output distribution, i.e. $100^{\circ}<\theta_{1}<250^{\circ}$  with a step of $7.5^{\circ}$, and $90^{\circ}<\theta_{2}<380^{\circ}$ with a step of $14.5^{\circ}$. The optimal angle combination for the D3-A6-C4 register is $\theta_{0}=90^{\circ}$, $\theta_{1}=212.5^{\circ}$, and $\theta_{2}=104.5^{\circ}$ (Fig.~\ref{3q}). As for the two-qubit case, we estimated errors for probability amplitudes by performing a statistic of $100$ measurements and by computing the differences of amplitudes between symmetric states. 

The experimental evidence of our ability to fine-tune the rotation angles for up to three qubits at a hardware level to achieve a specific output distribution demonstrates that the hardware and the experimental variational approach used here promise the possibility to legitimately address a more compelling quantum circuit, such as the GCI quantum circuit.
\begin{figure}[h]
	\centering
	\includegraphics[width=\columnwidth]{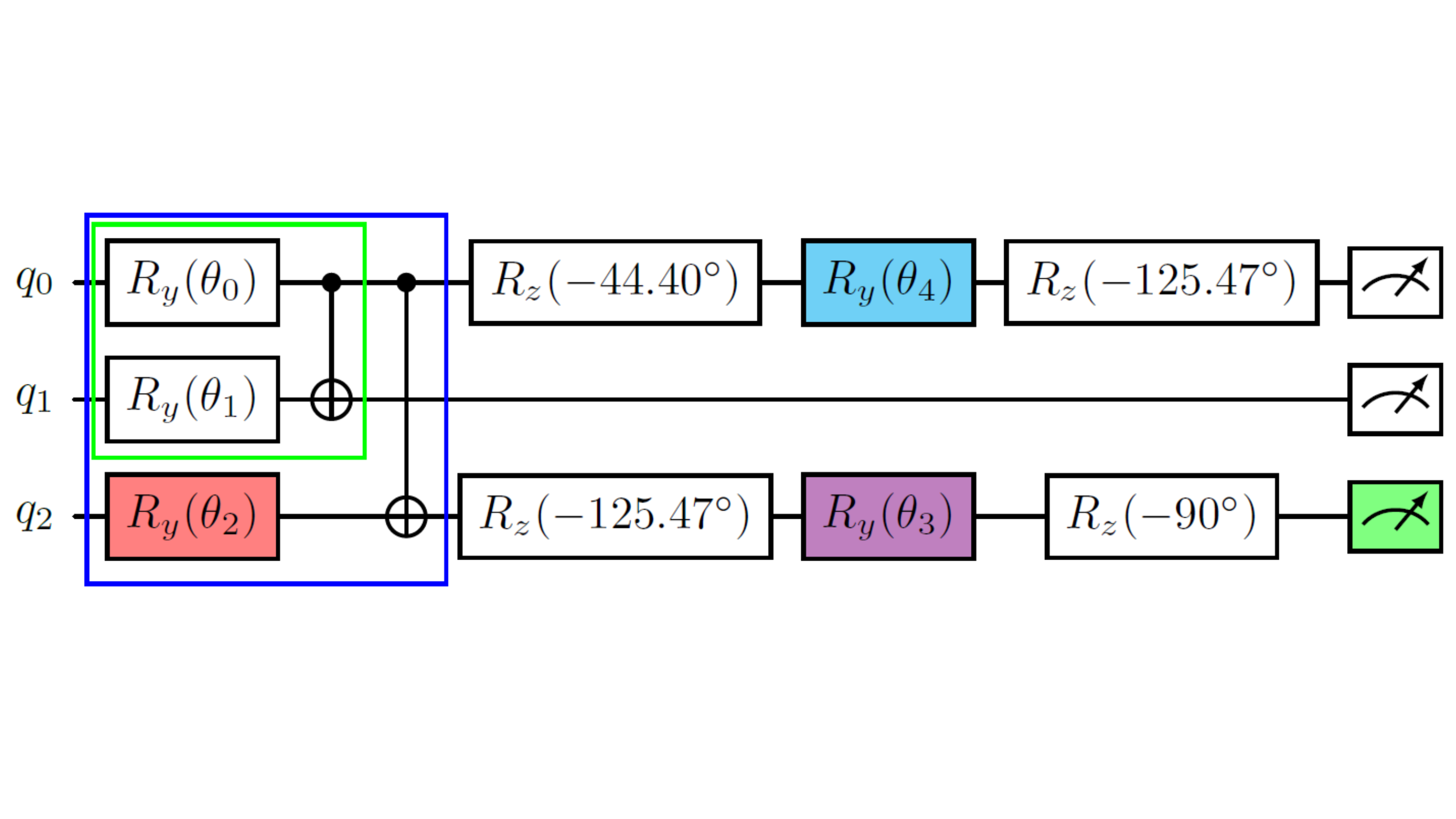}
\caption{Transpiled Gaussian Conditional Independence model quantum circuit. The first CNOT is here decomposed in terms of Hadamard and CZ gates, and includes a counter-phase Z gate with an angle of $-135^{\circ}$ to counteract CZ depolarizing errors, as discussed in Supplementary Material. The green box represents the 2-qubit normal distribution loading, while the blue box replaces the blue one in Fig.~\ref{TC}.}
\label{AssetAlg}
\end{figure}
\begin{table*}[t!]
	\centering
	\caption{GCI circuit outputs as a function of $\theta_{3}$ and $\theta_{4}$ at fixed $\theta_{2}=90^{\circ}$. In green, the use-case circuit output. In each plot is depicted the Probability Amplitude for each state in $\left\{|000\rangle,|001\rangle, |010\rangle, |011\rangle, |100\rangle, |101\rangle, |110\rangle, |111\rangle\right\}$ basis of D3-A6-C4 three-qubit register.}
	\includegraphics[width=\textwidth]{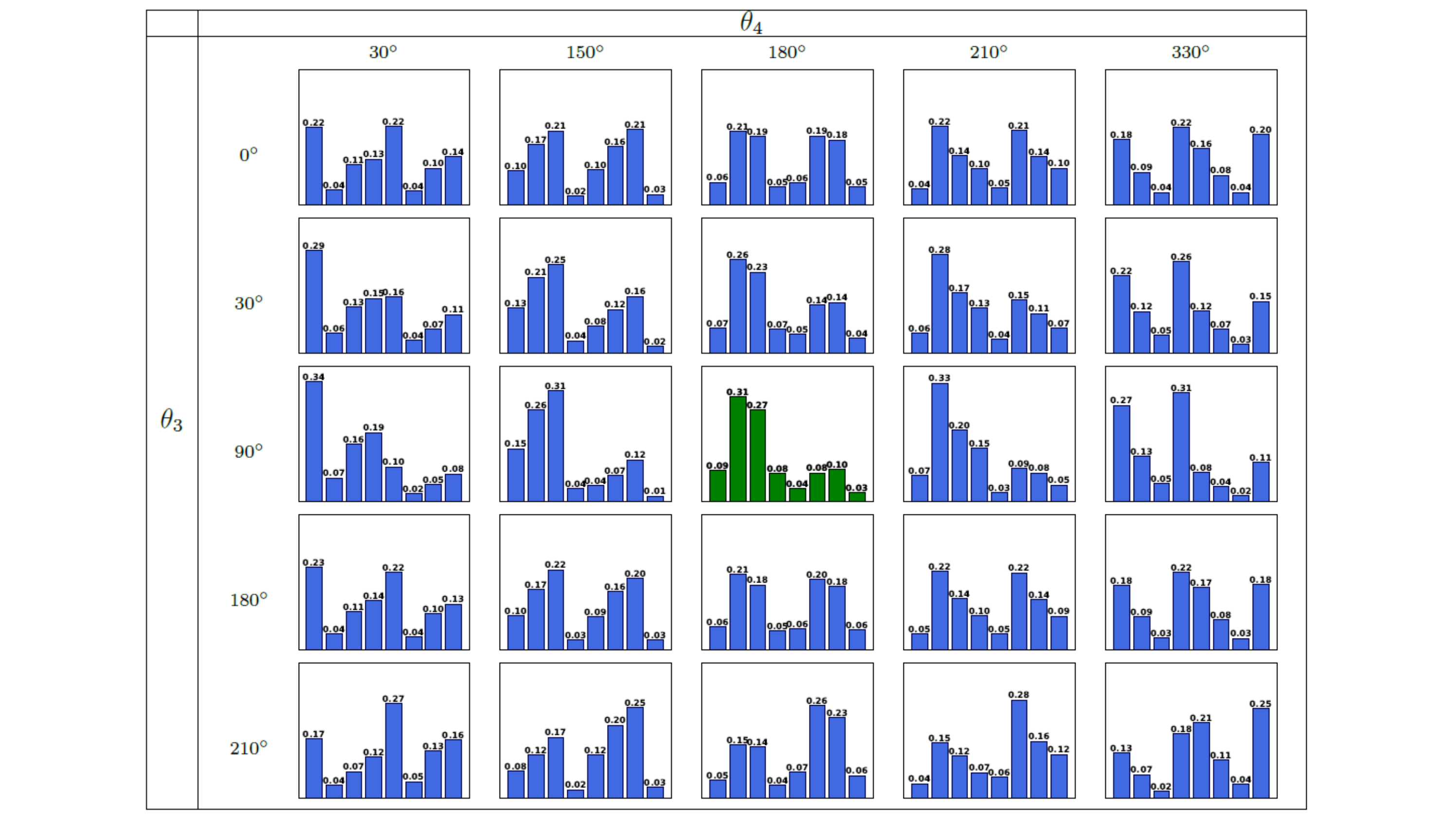}
	\label{T3vsT4}
\end{table*}
\begin{figure*}[t]   
\centering
\includegraphics[width=0.68\textwidth]{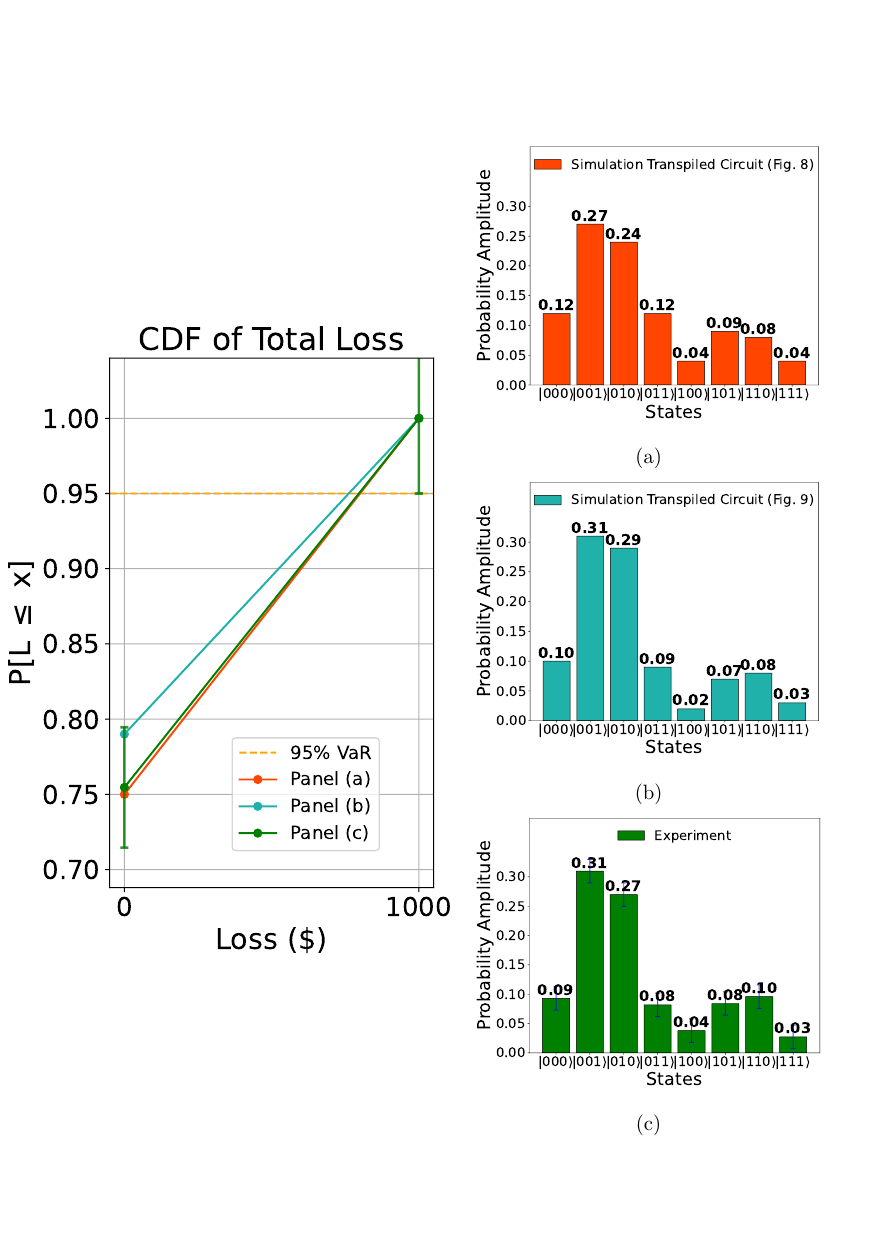}
\caption{Cumulative Distribution Function (CDF) of total losses for a loss $L=1000\$$ and a VaR of $95\%$, calculated from the probability distribution output of the GCI model in (a) (simulation of the transpiled quantum circuit in Fig.~\ref{TC}), in (b) (simulation of the transpiled quantum circuit in Fig.~\ref{AssetAlg}), and in (c) (experimental output of the same transpiled circuit). }
\label{CDFComparison}
\end{figure*}

\subsection{GCI quantum circuit}
\label{sec:results-GCIQC}

Our hardware-tested GCI circuit instantiates the one-risk-factor case for one asset ($k=1$ in Eq.~\ref{eq:gci-pd}): a latent $Z\sim\mathcal{N}(0,1)$ perturbs the asset’s default probability around $p^0_1=0.25$ with sensitivity $\rho_1=0.027$. The initial algorithm, introduced and shown in Fig.~\ref{fig:linrot-circuit}, requires three qubits and comprises a 2-qubit Gaussian sampling algorithm (green box) and one asset preparation and rotation gate, represented by a third qubit. This quantum circuit includes a controlled-Y (CY gate) that is prevented by the connectivity of the hardware. As follows, we detail the steps performed to implement the proposed quantum algorithm. \\ \textbf{Step 1: } we first transpiled the circuit using Qiskit’s preset pass manager, configured for hardware-aware optimization on the target backend. This pipeline performs comprehensive algebraic and structural simplifications (e.g., commutation, cancellation, and resynthesis) while respecting the backend’s basis gates and coupling constraints. For both initial qubit assignment and subsequent routing, we employed the SABRE (SWAP-Based Bidirectional heuristic search algorithm)~\cite{Niu2020}, which prioritizes mappings that reduce SWAP insertion and overall circuit depth on constrained topologies. Gate translation utilized synthesis-based methods with an approximation budget to allow slightly inexact unitary realizations in exchange for fewer two-qubit gates and shorter depths —an advantageous trade-off for NISQ hardware where fidelity is limited by circuit length. The final GCI transpiled quantum circuit is reported in Fig.~\ref{TC}. 

\textbf{Step 2}: we implemented the transpiled quantum circuit. Initially, the implementation of this quantum circuit on the machine led to unsatisfactory results, with significant deviations from the simulated output.

\textbf{Step 3}: we systematically simulated and experimentally implemented on the machine the transpiled circuit layer-by-layer. By systematically investigating the output of each quantum sub-circuit's layer through both the Qiskit simulator and the actual hardware, the output of the sub-circuit in the blue box of Fig.~\ref{TC} results in a uniform three-qubit distribution. The investigation detailed in Sec.~\ref{sec:results-3qnorm_dist} on the generation of a three-qubit normal distribution by exploiting the quantum circuit in Fig.~\ref{ThreeGauss} allowed us not only to understand for which rotation angles one can prepare a standard normal three-qubit distribution, but also to identify for which angles we were able to implement a uniform three-qubit distribution. In fact, Fig. S12 of the Supplementary Material shows that there are specific angles regions where a three-qubit uniform distribution can be eventually recovered. 

\textbf{Step 4}: we have opted for replacing the SABRE-based transpiled sub-circuit in the blue box with the one in Fig.~\ref{ThreeGauss}, achieving the configuration presented in Fig.~\ref{AssetAlg}. Here, the $R_y$ rotation angles $\theta_0$ and $\theta_1$ (green box) control the shape of the two-qubit distribution that represents the financial variable $z$. This sub-circuit must prepare a standard normal distribution, as previously discussed. As for the $R_y$ rotation angle $\theta_2$, conditioned by the angles $\theta_1$ and $\theta_0$, it controls on one hand the shape of the three-qubit uniform distribution that replaces the SABRE-based transpiled sub-circuit in the blue box of Fig.~\ref{TC}. On the other hand, it controls the asset preparation, and in part accounts for the transpiled realization, directly at the hardware level, of the GCI quantum circuit, together with the remaining $R_y$ rotation angles $\theta_3$ and $\theta_4$. In other words, 
\(R_Y(\theta_{2})\), \(R_Y(\theta_{3})\) and \(R_Y(\theta_{4})\) are initially derived from the offset and slope of the linearized probability of default in the noiseless quantum circuit model by the Adam and SABRE optimizers, but after transpilation and optimization, the circuit no longer preserves a direct correspondence between these angles and the original offset/slope parameters. Therefore, in the final hardware-ready transpiled quantum circuit, they should be regarded as tunable hyperparameters whose values are specific to the instance considered here. 

\textbf{Step 5}: having identified a reasonable transpilation of the initial GCI quantum circuit directly at the hardware level, the role of the rotation angles $\theta_2$, $\theta_3$ and $\theta_4$ changes. As tunable hyperparameters, we propose here to use the pulse-level variational approach used for the optimization of the Gaussian distribution, for the search of the optimal rotation angles in the final hardware-ready transpilation of the GCI quantum circuit. 

Specifically, we exploited qubits D3 and A6 to load the financial variable $z$ as a standard normal distribution. We recall that the optimal combination of rotation angles that gives a standard normal distribution for this pair is $\theta_{0}=90^{\circ}$ on D3 and $\theta_{1}=224^{\circ}$ on A6. C4 plays the role of the asset. Remarkably, the rotation angle $\theta_2$, combined with $\theta_{0}$ on D3 and $\theta_{1}$ on A6, allows to obtain the expected three-qubit uniform distribution at the blue box sub-circuit layer. We performed experiments by changing $\theta_{3,4}$ angles in the interval $\left[0^{\circ},360^{\circ}\right]$ 
to adapt the algorithm efficiency to the hardware, while keeping the asset angle rotation $90^\circ$ fixed $R_{Y}\left(\theta_{2}\right)$ (red square in Fig.~\ref{AssetAlg}).  In Tab.~\ref{T3vsT4}, we collect the experimental output of the GCI hardware-ready transpiled quantum circuit in Fig.~\ref{AssetAlg} as a function of the $\theta_{3,4}$ rotation angles. 

As done for the two- and three-qubit distribution preparations, we here discuss on the dependence of the output distributions on the varied angles. First of all, the experimental output resembles a bi-modal distribution in most cases. $R_{Y}\left(\theta_{3}\right)$ (purple square in Fig.~\ref{AssetAlg}) affects the height of the left or right gaussian with a periodicity of $90^{\circ}$, while $R_{Y}\left(\theta_{4}\right)$ (blue square in Fig.~\ref{AssetAlg}) acts on the single gaussian symmetry, with optimal behavior obtained for $R_{Y}\left(\theta_{4}\right) \in \left[150^{\circ},210^{\circ}\right]$. 
The profile that is most similar to the simulated output of the quantum circuit transpiled using Qiskit's preset pass manager (Fig.~\ref{CDFComparison} (a)) is highlighted in green and obtained for $\theta_{3}=90^{\circ}$ and $\theta_{4}=180^{\circ}$. 
As for the asset angle $R_{Y}\left(\theta_{2}\right)$ (red square in Fig.~\ref{AssetAlg}), in Tab. 2  of the Supplementary Material, we show that it is the experimental parameter that controls the symmetry between the two Gaussians at optimal $\{\theta_3,\theta_4\}=\{90^{\circ},180^{\circ}\}$. 

Finally, to assess the impact of the quantum implemented GCI on a relevant outcome for credit analysis we have used these data to calculate classically the CDF as follows. The pre-processing stage decomposes each measured bit string into two logical components: the leftmost bits encode default indicators, while the rightmost bits represent the latent variable \( Z \). For every observed outcome with an associated probability, the rightmost bits are converted to an integer index, allowing the marginal probability \( P(Z=z) \) to be accumulated. The default bits determine which debtors have defaulted; whenever a default bit is equal to one, the corresponding probability contributes to the marginal default probability, and the LGD value is added to the total loss for that scenario. After iterating over all outcomes, the procedure constructs (i) arrays of scenario losses and (ii) their probabilities, from which the expected loss is computed. Identical loss values are collapsed to form a discrete probability density function (e.g., the PDF) and its cumulative distribution function (e.g., the CDF). In Fig.~\ref{CDFComparison}, we compare the CDF for the simulated transpiled quantum circuit with Qiskit's preset pass manager (Fig.~\ref{TC}), the simulated hardware-ready transpiled quantum circuit (Fig.~\ref{AssetAlg}), and the experimental implementation on the hardware of the latter. The circuits' outputs are reported in panels (a), (b), and (c), respectively.

The hardware-ready transpilation on the machine, affected by a readout error of about $\sim 5\%$, produces a CDF (Panel c)  with a discrepancy of $\sim0.5\%$, with respect to the simulation of the Transpiled circuit (Panel a), indeed showing a strong agreement with the expected CDF : $P(L \le 0) \approx 0.75$ and $P(L \le 1000)=1.0$. The main result is therefore not the recovery of the $95\%$ VaR itself, since the VaR is classically determined from the reference distribution, but rather the fact that the quantum algorithm produces the same distribution from which the VaR is obtained. Specifically, the Hellinger fidelity~\cite{Hellinger1909} between the experimental distribution (Fig.~\ref{CDFComparison} (c)) and the distribution simulated using the transpiled quantum circuit in Fig.~\ref{TC} (panel~\ref{CDFComparison} (a)) is $F_{H}=(98.9\pm0.3)\%$. Therefore, the quantum and classical outputs leads the same distributional description and, consequently, to the same risk conclusion for this scenario.  This confirms the validity of our experimental approach, which relies entirely on the performance of the available NISQ hardware, taken into account through the variational hyper-tuning performed directly on the machine. More generally, this result highlights that transpilation, while effectively re-assigning and re-combining unitary rotations to address connectivity constraints and circuit depth, does not account for the sensitivity of an NISQ device to specific noise sources, even though such information is essential in the NISQ era.

\section{Discussion}
\label{sec:discussions}
The development of quantum algorithms for finance stems as an impactful topic, which on one hand requires to design algorithms that provide a measurable improvement compared to their classical counterpart, and on the other there is a strong requirement of compliance with current state-of-the-art quantum hardware to experimentally demonstrate their feasibility. \\In our knowledge, one of the most common approach in algorithm development implicitly rely on the fact that quantum processors behave ideally without noise, or fault-tolerantly~\cite{Pan2021}. Although there is a strong effort and interest in achieving fault-tolerant quantum computing in the near term, with a substantial research on Quantum Error Correction from both software and hardware point of views~\cite{Campbell2017}, current NISQ quantum processors are not yet capable of sustaining large-scale problems. \\The development of quantum algorithms in the NISQ-era by noise-modeling current quantum machines~\cite{Clinton2021,Bordoni2025,Du2026}, promises to anticipate the outcome before running on the real processor. In superconducting quantum processors, for example, one can mention bit-flip, phase-flip and decoherence errors as the most important noise sources~\cite{Ghosh2012}. Although available circuit simulators include reasonable noise-models for superconducting systems, these do not necessarily apply to every quantum machine design, electrodynamics and circuital characteristics. In order to efficiently develop quantum algorithms, it is still required a strong effort to model most, if not all, noise sources of the device, since they may not be limited to the ones aforementioned. In this work, we have, in fact, relied on off-line noiseless quantum circuit simulation for a first optimization of rotation angles, which provided us a starting point for the implementation of a GCI model on a NISQ processor. \\The findings of this work have highlighted three important noise sources that should be tackled and eventually included in a noise-model: readout noise, phase errors and depolarizing errors. As discussed in Sec.~\ref{sec:results}, we investigated on the impact of fluctuations on the output states discrimination due to SPAM errors, and provided a statistical analysis to evaluate a maximum error to consider when comparing simulations and experimental results. Depolarizing errors have been addressed by including a counter-phase Z gate in the CNOT implementation, as detailed in Supplementary Material. As for phase-errors related to undershoot or overshoot of control pulses used to implement $R_y$ rotations, these have been accounted for by the variational technique deployed in this work. \\
An alternative approach would rely on approximate noise models by using existing circuit simulators, by counteracting remaining noise sources with error mitigation techniques.  Mitigating errors in real devices is also not trivial~\cite{Cai2023}. Efficient error mitigation requires to detect and identify what noise sources are mostly impacting the machine, and consistently apply a useful mitigation strategy. Readout error mitigation, for example, allows to optimize for errors in readout state discrimination, and eventually address bit-flip errors~\cite{ahmad2024_mit}. Dynamical decoupling~\cite{Viola1998} allows to mitigate idling errors by continuously executing sequences of single-qubit operations on idle qubits that collectively function as identity gates, thereby suppressing environmental noise and relaxation. Depolarizing error mitigation, instead, involves techniques to counteract decoherence. To address depolarizing errors, common methods include Zero-Noise Extrapolation (ZNE)~\cite{Cai2023}, a randomized compiling to convert coherent errors into depolarizing noise, and estimating global depolarizing rates via partner circuits to rescale results. In principle, the identification of the best error mitigation technique relies on a systematic investigation and experimental characterization, thus not necessarily providing a more efficient solution if compared to variational approaches. \\Most importantly, the study developed here provided a useful playground to identify specific noise sources that, on one hand, will be modeled and taken into account in a circuit simulator to optimize rotation angles off-line prior to the experimental realization on the hardware; on the other, the direct access to the quantum processor allowed us to devise an experimental variational approach able to counteract such effects directly in-situ, while still being able to successfully implement the GCI quantum circuit. Remarkably, the hardware-aware variational approach here proposed allowed to counteract for systematic errors on the experimental control electronics: pulsed signals phase-errors, undershoot or overshoot of pulsed signal amplitudes, delays in the control lines setup, which are not entirely compensated by calibration and optimization mechanisms of single- and two-qubit gates rotations and would have required tailored or multiple error mitigation techniques. In a way, we exploited the implementation of proof-of-concept quantum algorithms on real NISQ devices as a tool to study noise and errors mechanisms, and to include in the optimization task the real nature of the hardware. \\Although the Quantum circuit-based adaptation approach applies in principle to every superconducting quantum processor, and to quantum algorithms designed for different applications, one must be aware that such a hardware-aware variational approach is not hardware-agnostic. Such an optimization depends on the processor, and in case of extended tasks involving more random variables and more qubits, as predicted in Sec.~\ref{sec:methods}, the optimization would need to be repeated for each variable in play. For example, we expect a dependence of the performance on the quantum circuit scales. Let us consider an extension from a one-risk-factor/one-asset quantum circuit to a quantum circuit that encodes two risk factors and two assets. The Qiskit's preset pass manager transpilation requires several quantum gates and qubits that roughly double with the toy-model addressed in this work. We have measured readout fidelities increasing the number of qubits from $1$ to $4$, and reported the readout fidelity dependence as a function of the number of qubits in Fig. S8 of Supplementary Material. The readout fidelity scales linearly with the number of qubits, and we can project to six qubits. In a three-qubit register, the readout fidelity is $\sim88\%$, while in a six-qubit register, we expect a readout fidelity $\sim75\%$. For these numbers, state discrimination will likely be affected, and readout error mitigation will be necessary. As for what concerns the coherence impact, and considering the single- and two-qubit gate durations reported in Supplementary Material, the total duration of the Qiskit's preset pass-manager transpiled circuit implemented on hardware was $\sim 3\mu s$, and has been reduced by a factor 2 in the hardware-aware transpiled circuit ($\sim 1.6 \mu s$). For a two-risk-factor/two-asset configuration, the Qiskit's preset pass manager transpiled algorithm would require a circuit of $31$ layers of gates, with an estimated total duration of $\sim 5\mu s$. The major contribution to the running time is due to the number of $CNOT$ gates. Although we require experimentally implementing the algorithm to evaluate the scaling impact of the hardware-aware transpilation, it is fair to expect to achieve a reduction in time of a factor of $2$ as in the one-risk-factor/one-asset scenario. By comparing such durations with the lowest coherence time of the register, i.e. $T_{2}^{*}=20 \mu s$, the three-qubit original transpiled circuit duration is $15\%$ of $T_{2}^{*}$, and the six-qubit circuit projection should be the $25\%$ of the coherence time. Therefore, we do not expect that coherence time is a limit for scaling to two-risk-factors/two-assets.\\ Finally, we have also reported a significant dependence on the qubits chosen in a register, therefore confirming the critical role played by the knowledge of the hardware to develop quantum algorithms in the NISQ era.

\section{Conclusion}
\label{sec:conclusion}

In this work, we have provided a systematic experimental investigation of how variational quantum circuits can model distributions relevant to CRA on a superconducting quantum computer: two- and three-qubit Gaussian distributions and the GCI. 

We have examined how sQPU connectivity, the different circuit nature of superconducting qubits in quantum registers, and the quantum circuit transpilation itself (e.g., entangler placement and single-qubit rotation settings) influence the desired distribution, encoded in the probability amplitudes of the prepared quantum states. A remarkable result is that the rotation angles required to realize a normal-like histogram are not generic but depend on the specific hardware characteristics of the device. We have experimentally analyzed the impact of single-qubit rotation angles on the shape of probability amplitudes distributions encoded in both two- and three-qubits registers.  
This allowed us not only to identify the optimal angle configuration but also, and most importantly, to determine the experimental confidence error threshold in a typical NISQ device by implementing hundreds of experimental runs. 

The in-situ variational approach introduced here has also been used to optimize the GCI quantum circuit output for fixed problem parameters, taken here as an example, obtaining excellent agreement with the classical counterpart. In a sense, we have performed a hardware-aware transpilation of the quantum circuits directly at the pulse level for this use case, allowing us to capture the realistic nature of the hardware, down to the level of the individual qubits' performances and connectivities. The validity of our approach has been confirmed by the calculation of the CDF for a specific use case, obtained by using the experimental data derived from an \emph{ad-hoc} transpilation of the quantum circuit, implicitly taking into account the errors of the machine, and by simulating the same quantum circuit with a noiseless simulator. Such dependence is critical in the NISQ era and emphasizes the requirement for a clear understanding of quantum computing in all its parts to achieve quantum utility in the near term: from algorithm coding to implementation through analog pulses, which allowed us to construct a more efficient quantum circuit in terms of gate depth. 

Our investigation demonstrates that the readout fidelities achieved on the processor already offer satisfying results, since the Hellinger fidelities, used as a metric to quantify the deviation between the quantum and classical simulated outputs for the use-cases analyzed here, are remarkably close to the state-of-the-art values and larger than more commonly reported in literature~\cite{Schulze2021,ahmad2024_mit}. We specify that no readout error mitigation technique has been applied, nor have we integrated quantum-noise limited cryogenic amplifiers at the hardware stage. However, since we expect that the readout fidelity tends to decrease with increasing number of qubits in the registers (see Fig.8 in Supplementary Materials), we envision including both into our quantum algorithms implementation pipeline in the near term~\cite{levochkina2024investigating,ahmad2024_mit,Ahmad2025_magic}. Moreover, a systematic implementation of variational quantum algorithms and circuits, not necessarily limited to the ones explored in this work, and that will eventually involve different qubits in the chip, will likely allow to uniquely define the deviation of rotation angles from the noiseless case for a fixed register. This will allow to obtain reproducible outputs on different registers of the quantum processor, once the optimized angles have been determined for each register. Moving to the envisioned generic approach in NISQ devices, the possibility to exploit knowledge on hardware errors and noise from quantum circuits implementation is the key to develop a tailored noise-model, which will be used to simulate quantum circuits, especially when a large number of qubits and parameters will be needed.

More and more research centers and industries will rely on proprietary superconducting hardware designs in the near term, with their own geometries, connectivity, characteristic circuits, and electronic parameters for quantum gate implementation. We believe this work provides a guide towards the hardware-aware implementation of other proof-of-concept quantum algorithms and motivates the use of quantum compilers that adapt circuit structure and parameterization to the specific QPU, rather than relying on a fixed, hardware-agnostic configuration. 

\section*{Fundings}
The work was supported by the PNRR MUR project CN00000013-ICSC, the PNRR MUR project PE0000023-NQSTI, the Pathfinder EIC 2023 project "FERROMON-Ferrotransmons and Ferrogatemons for Scalable Superconducting Quantum Computers" and the Project PRIN 2022-Advanced Control and Readout of scalable Superconducting NISQ Architectures (SuperNISQ)-CUP E53D23001910006.

\section*{Acknowledgment}
H.G.A., D.M., and F.T. thank the SUPERQUMAP project (COST Action CA21144) and the support of Hany Ali and Alessandro Bruno from Quantware. We also acknowledge support from Johannes Hermann and Christian Junger. We also thank Emanuele Dri from LINKS Foundation for providing useful insights in the initial stage of the project.

\section*{Data Availability Statement}

The raw data that support the findings of this study are publicly available via Zenodo Platform at\\ https://doi.org/10.5281/zenodo.19402577.

\section{Conflict of Interest statement}

The authors declare no conflicts of interest.

\bibliographystyle{IEEEtran}

\begin{thebibliography}{10}
		\providecommand{\url}[1]{#1}
		\csname url@samestyle\endcsname
		\providecommand{\newblock}{\relax}
		\providecommand{\bibinfo}[2]{#2}
		\providecommand{\BIBentrySTDinterwordspacing}{\spaceskip=0pt\relax}
		\providecommand{\BIBentryALTinterwordstretchfactor}{4}
		\providecommand{\BIBentryALTinterwordspacing}{\spaceskip=\fontdimen2\font plus
			\BIBentryALTinterwordstretchfactor\fontdimen3\font minus
			\fontdimen4\font\relax}
		\providecommand{\BIBforeignlanguage}[2]{{%
				\expandafter\ifx\csname l@#1\endcsname\relax
				\typeout{** WARNING: IEEEtran.bst: No hyphenation pattern has been}%
				\typeout{** loaded for the language `#1'. Using the pattern for}%
				\typeout{** the default language instead.}%
				\else
				\language=\csname l@#1\endcsname
				\fi
				#2}}
		\providecommand{\BIBdecl}{\relax}
		\BIBdecl
		
		\bibitem{Egger2021}
		\BIBentryALTinterwordspacing
		D.~J. Egger, R.~Garcia~Gutierrez, J.~C. Mestre, and S.~Woerner, ``Credit risk
		analysis using quantum computers,'' \emph{IEEE Transactions on Computers},
		vol.~70, no.~12, p. 2136–2145, Dec 2021. [Online]. Available:
		\url{http://dx.doi.org/10.1109/TC.2020.3038063}
		\BIBentrySTDinterwordspacing
		
		\bibitem{Dri2023}
		\BIBentryALTinterwordspacing
		E.~Dri, A.~Aita, E.~Giusto, D.~Ricossa, D.~Corbelletto, B.~Montrucchio, and
		R.~Ugoccioni, ``A more general quantum credit risk analysis framework,''
		\emph{Entropy}, vol.~25, no.~4, 2023. [Online]. Available:
		\url{https://www.mdpi.com/1099-4300/25/4/593}
		\BIBentrySTDinterwordspacing
		\bibitem{veronelli2025implementingcreditriskanalysis}
		\BIBentryALTinterwordspacing
		D.~Veronelli, F.~Cibrario, E.~Dri, V.~Zaffaroni, G.~Ranieri, D.~Corbelletto,
		and B.~Montrucchio, ``Implementing credit risk analysis with quantum singular
		value transformation,'' 2025. [Online]. Available:
		\url{https://arxiv.org/abs/2507.19206}
		\BIBentrySTDinterwordspacing
		
		\bibitem{Montanaro2015}
		\BIBentryALTinterwordspacing
		A.~Montanaro, ``Quantum speedup of Monte Carlo methods,'' \emph{Proceedings of
			the Royal Society A: Mathematical, Physical and Engineering Sciences}, vol.
		471, no. 2181, p. 20150301, Sep 2015. [Online]. Available:
		\url{http://dx.doi.org/10.1098/rspa.2015.0301}
		\BIBentrySTDinterwordspacing
		
		\bibitem{Chen2023}
		\BIBentryALTinterwordspacing
		S.~Chen, J.~Cotler, H.-Y. Huang, and J.~Li, ``The complexity of NISQ,''
		\emph{Nature Communications}, vol.~14, no.~1, p. 6001, Sep 2023. [Online].
		Available: \url{https://doi.org/10.1038/s41467-023-41217-6}
		\BIBentrySTDinterwordspacing
		
		\bibitem{Devoret2013}
		\BIBentryALTinterwordspacing
		M.~H. Devoret and R.~J. Schoelkopf, ``Superconducting circuits for quantum
		information: An outlook,'' \emph{Science}, vol. 339, no. 6124, pp.
		1169--1174, 2013. [Online]. Available:
		\url{https://www.science.org/doi/abs/10.1126/science.1231930}
		\BIBentrySTDinterwordspacing
		
		\bibitem{Salvoni2021}
		D.~Salvoni, L.~Parlato, M.~Ejrnaes, F.~Mattioli, A.~Gaggero, F.~Martini,
		G.~Ausanio, D.~Massarotti, D.~Montemurro, H.~G. Ahmad, L.~di~Palma,
		F.~Tafuri, R.~Cristiano, and G.~P. Pepe, ``Demonstration of single photon
		detection in amorphous molybdenum silicide / aluminium superconducting
		nanostrip,'' \emph{IEEE Instrumentation \& Measurement Magazine}, vol.~24,
		no.~5, pp. 69--74, 2021.
		
		\bibitem{Mastrovito2024}
		P.~Mastrovito, H.~G. Ahmad, A.~Porzio, M.~Esposito, D.~Massarotti, and
		F.~Tafuri, ``{On-chip microwave coherent source with in-situ control of the
			photon number distribution},'' \emph{Commun. Phys.}, vol.~8, no.~1, p. 295,
		2025.
		
		\bibitem{Rasmussen2021}
		\BIBentryALTinterwordspacing
		S.~Rasmussen, K.~Christensen, S.~Pedersen, L.~Kristensen, T.~B\ae{}kkegaard,
		N.~Loft, and N.~Zinner, ``Superconducting circuit companion---an introduction
		with worked examples,'' \emph{PRX Quantum}, vol.~2, p. 040204, Dec 2021.
		[Online]. Available:
		\url{https://link.aps.org/doi/10.1103/PRXQuantum.2.040204}
		\BIBentrySTDinterwordspacing
		
		\bibitem{Krantz2019}
		\BIBentryALTinterwordspacing
		P.~Krantz, M.~Kjaergaard, F.~Yan, T.~P. Orlando, S.~Gustavsson, and W.~D.
		Oliver, ``A quantum engineer's guide to superconducting qubits,''
		\emph{Applied Physics Reviews}, vol.~6, no.~2, p. 021318, 06 2019. [Online].
		Available: \url{https://doi.org/10.1063/1.5089550}
		\BIBentrySTDinterwordspacing
		
		\bibitem{ahmad2023}
		H.~G. Ahmad, C.~Jordan, R.~van~den Boogaart, D.~Waardenburg, C.~Zachariadis,
		P.~Mastrovito, A.~L. Georgiev, D.~Montemurro, G.~P. Pepe, M.~Arthers
		\emph{et~al.}, ``Investigating the individual performances of coupled
		superconducting transmon qubits,'' \emph{Condensed Matter}, vol.~8, no.~1,
		p.~29, 2023.
		
		\bibitem{stasino2025implementation}
		V.~Stasino, P.~Mastrovito, C.~Cosenza, A.~Levochkina, M.~Esposito,
		D.~Montemurro, G.~P. Pepe, A.~Bruno, F.~Tafuri, D.~Massarotti \emph{et~al.},
		``Implementation and readout of maximally entangled two-qubit gates quantum
		circuits in a superconducting quantum processor,'' \emph{Journal of
			Superconductivity and Novel Magnetism}, vol.~38, no.~2, p. 129, 2025.
		
		\bibitem{Beck2024}
		\BIBentryALTinterwordspacing
		T.~Beck, A.~Baroni, R.~Bennink, G.~Buchs, E.~A.~C. Pérez, M.~Eisenbach, R.~F.
		{da Silva}, M.~G. Meena, K.~Gottiparthi, P.~Groszkowski, T.~S. Humble,
		R.~Landfield, K.~Maheshwari, S.~Oral, M.~A. Sandoval, A.~Shehata, I.-S. Suh,
		and C.~Zimmer, ``Integrating quantum computing resources into scientific
		{HPC} ecosystems,'' \emph{Future Generation Computer Systems}, vol. 161, pp.
		11--25, 2024. [Online]. Available:
		\url{https://www.sciencedirect.com/science/article/pii/S0167739X24003583}
		\BIBentrySTDinterwordspacing
		
		\bibitem{Mansfield2025}
		\BIBentryALTinterwordspacing
		E.~Mansfield, S.~Seegerer, P.~Vesanen, J.~Echavarria, B.~Mete, M.~N. Farooqi,
		and L.~Schulz, ``First practical experiences integrating quantum computers
		with HPC resources: A case study with a 20-qubit superconducting quantum
		computer,'' 2025. [Online]. Available: \url{https://arxiv.org/abs/2509.12949}
		\BIBentrySTDinterwordspacing
		
		\bibitem{Ahmad2022_ferro}
		\BIBentryALTinterwordspacing
		H.~G. Ahmad, V.~Brosco, A.~Miano, L.~Di~Palma, M.~Arzeo, D.~Montemurro,
		P.~Lucignano, G.~P. Pepe, F.~Tafuri, R.~Fazio, and D.~Massarotti, ``Hybrid
		ferromagnetic transmon qubit: Circuit design, feasibility, and detection
		protocols for magnetic fluctuations,'' \emph{Phys. Rev. B}, vol. 105, p.
		214522, Jun 2022. [Online]. Available:
		\url{https://link.aps.org/doi/10.1103/PhysRevB.105.214522}
		\BIBentrySTDinterwordspacing
		
		\bibitem{Hu2024}
		\BIBentryALTinterwordspacing
		F.~Hu, S.~A. Khan, N.~T. Bronn, G.~Angelatos, G.~E. Rowlands, G.~J. Ribeill,
		and H.~E. T{\"u}reci, ``Overcoming the coherence time barrier in quantum
		machine learning on temporal data,'' \emph{Nature Communications}, vol.~15,
		no.~1, p. 7491, Aug 2024. [Online]. Available:
		\url{https://doi.org/10.1038/s41467-024-51162-7}
		\BIBentrySTDinterwordspacing
		
		\bibitem{Tuokkola2025}
		\BIBentryALTinterwordspacing
		M.~Tuokkola, Y.~Sunada, H.~Kivij{\"a}rvi, J.~Albanese, L.~Gr{\"o}nberg, J.-P.
		Kaikkonen, V.~Vesterinen, J.~Govenius, and M.~M{\"o}tt{\"o}nen, ``Methods to
		achieve near-millisecond energy relaxation and dephasing times for a
		superconducting transmon qubit,'' \emph{Nature Communications}, vol.~16,
		no.~1, p. 5421, Jul 2025. [Online]. Available:
		\url{https://doi.org/10.1038/s41467-025-61126-0}
		\BIBentrySTDinterwordspacing
		
		\bibitem{Preskill1998}
		J.~Preskill, ``Fault-tolerant quantum computation,'' \emph{Introduction to
			quantum computation and information}, vol. 213, 1998.
		
		\bibitem{Stamatopoulos2020}
		\BIBentryALTinterwordspacing
		N.~Stamatopoulos, D.~J. Egger, Y.~Sun, C.~Zoufal, R.~Iten, N.~Shen, and
		S.~Woerner, ``Option {P}ricing using {Q}uantum {C}omputers,''
		\emph{{Quantum}}, vol.~4, p. 291, Jul 2020. [Online]. Available:
		\url{https://doi.org/10.22331/q-2020-07-06-291}
		\BIBentrySTDinterwordspacing
		
		\bibitem{chakrabarti2021}
		\BIBentryALTinterwordspacing
		S.~Chakrabarti, R.~Krishnakumar, G.~Mazzola, N.~Stamatopoulos, S.~Woerner, and
		W.~J. Zeng, ``A threshold for quantum advantage in derivative pricing,''
		\emph{Quantum}, vol.~5, p. 463, Jun 2021. [Online]. Available:
		\url{http://dx.doi.org/10.22331/q-2021-06-01-463}
		\BIBentrySTDinterwordspacing
		
		\bibitem{Cibrario2024}
		F.~Cibrario, O.~S. Golan, G.~Ranieri, E.~Dri, M.~Ippoliti, R.~Cohen, C.~Mattia,
		B.~Montrucchio, A.~Naveh, and D.~Corbelletto, ``Quantum amplitude loading for
		rainbow options pricing,'' in \emph{2024 IEEE International Conference on
			Quantum Computing and Engineering (QCE)}, vol.~01, 2024, pp. 211--220.
		
		\bibitem{cibrario2025autocallableoptionspricingintegrationbased}
		\BIBentryALTinterwordspacing
		F.~Cibrario, R.~Cohen, E.~Dri, C.~Mattia, O.~S. Golan, T.~Danzig, G.~Ranieri,
		H.~Rosemarin, D.~Corbelletto, A.~Naveh, and B.~Montrucchio, ``Autocallable
		options pricing with integration-based exponential amplitude loading,'' 2025.
		[Online]. Available: \url{https://arxiv.org/abs/2507.19039}
		\BIBentrySTDinterwordspacing
		
		\bibitem{doi:10.1142/S021902491550034X}
		\BIBentryALTinterwordspacing
		M.~RUTKOWSKI and S.~TARCA, ``Regulatory capital modeling for credit risk,''
		\emph{International Journal of Theoretical and Applied Finance}, vol.~18,
		no.~05, p. 1550034, 2015. [Online]. Available:
		\url{https://doi.org/10.1142/S021902491550034X}
		\BIBentrySTDinterwordspacing
		
		\bibitem{Vasicek2002}
		O.~A. Vasicek, ``The distribution of loan portfolio value,'' \emph{Risk}, Dec.
		2002, reprinted online by Risk.net.
		
		\bibitem{Gordy2003}
		M.~B. Gordy, ``A risk-factor model foundation for ratings-based bank capital
		rules,'' \emph{Journal of Financial Intermediation}, vol.~12, no.~3, pp.
		199--232, 2003.
         \bibitem{loaddistrubition}
		\BIBentryALTinterwordspacing
		Binoy Paine, Kalyan Dasgupta, ``Loading Probability Distributions in a Quantum circuit
        ,'' Available:
		\url{https://arxiv.org/abs/2208.13372}
		\BIBentrySTDinterwordspacing
		
		\bibitem{qgan}
		\BIBentryALTinterwordspacing
		C.~Zoufal, A.~Lucchi, and S.~Woerner, ``Quantum generative adversarial networks
		for learning and loading random distributions,'' \emph{npj Quantum
			Information}, vol.~5, 2019. [Online]. Available:
		\url{https://arxiv.org/pdf/1904.00043}
		\BIBentrySTDinterwordspacing
		
		\bibitem{Kingma2014AdamAM}
		\BIBentryALTinterwordspacing
		D.~P. Kingma and J.~Ba, ``Adam: A method for stochastic optimization,''
		\emph{CoRR}, vol. abs/1412.6980, 2014. [Online]. Available:
		\url{https://api.semanticscholar.org/CorpusID:6628106}
		\BIBentrySTDinterwordspacing

        \bibitem{Ballarin2025}
		\BIBentryALTinterwordspacing
		M.~Ballarin, J. J.~ García-Ripoll, D.~Hayes, M.~Lubasch, ``Efficient quantum state preparation of multivariate functions using tensor networks,'', Arxiv, 2025. [Online]. Available:
		\url{https://arxiv.org/abs/2511.15674}
        
		\BIBentrySTDinterwordspacing
        \bibitem{Pellow2021}
		\BIBentryALTinterwordspacing
		A.~ Pellow-Jarman, I.~ Sinayskiy, A.~ Pillay, F.~ Petruccione, ``A comparison of various classical optimizers for a variational quantum linear solver,'', Quantum Inf Process, vol. 20, p. 202, 2021. [Online]. Available: \url{https://doi.org/10.1007/s11128-021-03140-x}
		\BIBentrySTDinterwordspacing

        \BIBentrySTDinterwordspacing
        \bibitem{Riste2020}
		\BIBentryALTinterwordspacing
		D.~Rist\`e, S.~Fallek, B.~Donovan, T.A.~Ohki, ``Microwave Techniques for Quantum Computers: State-of-the-Art Control Systems for Quantum Processors,'', IEEE Microwave Magazine, vol. 21, no. 8, p. 60-71, 2020. [Online]. Available: \url{https://doi.org/10.1109/MMM.2020.2993477}
		\BIBentrySTDinterwordspacing
		
		\bibitem{Negirnac2021}
		\BIBentryALTinterwordspacing
		V.~Neg\^{\i}rneac, H.~Ali, N.~Muthusubramanian, F.~Battistel, R.~Sagastizabal,
		M.~S. Moreira, J.~F. Marques, W.~J. Vlothuizen, M.~Beekman, C.~Zachariadis,
		N.~Haider, A.~Bruno, and L.~DiCarlo, ``High-fidelity controlled-$Z$ gate with
		maximal intermediate leakage operating at the speed limit in a
		superconducting quantum processor,'' \emph{Phys. Rev. Lett.}, vol. 126, p.
		220502, Jun 2021. [Online]. Available:
		\url{https://link.aps.org/doi/10.1103/PhysRevLett.126.220502}
		\BIBentrySTDinterwordspacing
		
		\bibitem{Ahmad2025_magic}
		\BIBentryALTinterwordspacing
		H.~G. Ahmad, G.~Esposito, V.~Stasino, J.~Odavic, C.~Cosenza, A.~Sarno,
		P.~Mastrovito, M.~Viscardi, S.~Cusumano, F.~Tafuri, D.~Massarotti, and
		A.~Hamma, ``Experimental demonstration of non-local magic in a
		superconducting quantum processor,'' 2025. [Online]. Available:
		\url{https://arxiv.org/abs/2511.15576}
		\BIBentrySTDinterwordspacing
		
		\bibitem{Quantify}
		\BIBentryALTinterwordspacing
		Quantify, ``Quantify python framework,'' quantify Documentation. Accessed:
		2025-12-13. [Online]. Available: \url{https://quantify-os.org/}
		\BIBentrySTDinterwordspacing
		
		\bibitem{Zvirtual}
		\BIBentryALTinterwordspacing
		Qblox, ``Numerically controlled oscillator,'' qblox Documentation 2025.10.0.
		Accessed: 2025-11-03. [Online]. Available:
		\url{https://docs.qblox.com/en/main/products/qBlox\_instruments/
			tutorials/QRM/nco\_control\_adv.html}
		\BIBentrySTDinterwordspacing
		
		\bibitem{McKay2017}
		\BIBentryALTinterwordspacing
		D.~C. McKay, C.~J. Wood, S.~Sheldon, J.~M. Chow, and J.~M. Gambetta,
		``Efficient {$Z$} gates for quantum computing,'' \emph{Phys. Rev. A},
		vol.~96, p. 022330, Aug 2017. [Online]. Available:
		\url{https://link.aps.org/doi/10.1103/PhysRevA.96.022330}
		\BIBentrySTDinterwordspacing

        \bibitem{Hellinger1909}
		\BIBentryALTinterwordspacing
		E.~Hellinger,
		``Neue begr{\"u}ndung der theorie quadratischer formen von unendlichvielen ver{\"a}nderlichen,'' \emph{Journal f{\"u}r die reine und angewandte Mathematik},
		vol.~1909, no. 136, p. 210-271, 1909. [Online]. Available:
		\url{https://doi.org/10.1515/crll.1909.136.210}
		\BIBentrySTDinterwordspacing
		
		\bibitem{Niu2020}
		S.~Niu, A.~Suau, G.~Staffelbach, and A.~Todri-Sanial, ``A hardware-aware
		heuristic for the qubit mapping problem in the nisq era,'' \emph{IEEE
			Transactions on Quantum Engineering}, vol.~1, pp. 1--14, 2020.

        \bibitem{Pan2021}
		\BIBentryALTinterwordspacing
		J.~Pan, ``Analyzing noise for quantum advantage,''
		\emph{Nature Computational Science}, vol.~1, no.~12, p. 776, Dec 2021. [Online].
		Available: \url{https://doi.org/10.1038/s43588-021-00178-w}
		\BIBentrySTDinterwordspacing

        \bibitem{Campbell2017}
		\BIBentryALTinterwordspacing
		E.~Campbell, B.~Terhal, C.~Vuillot, ``Roads towards fault-tolerant universal quantum computation,''
		\emph{Nature}, vol.~549, p. 172-179, 2017. [Online].
		Available: \url{https://doi.org/10.1038/nature23460}
		\BIBentrySTDinterwordspacing

        \bibitem{Clinton2021}
		\BIBentryALTinterwordspacing
		L.~Clinton, J.~Bausch, T.~Cubitt, ``Hamiltonian simulation algorithms for near-term quantum hardware,''
		\emph{Nature Communications}, vol.~12, p. 4989, 2021. [Online].
		Available: \url{https://doi.org/10.1038/s41467-021-25196-0}
		\BIBentrySTDinterwordspacing

        \bibitem{Bordoni2025}
		\BIBentryALTinterwordspacing
		S.~Bordoni, A.~Papaluca, P.~Buttarini, A.~Sopena, S.~Giagu, S.~Carrazza, ``Quantum noise modeling through reinforcement learning,''
		\emph{Quantum Sci. Technol.}, vol.~11, p. 015005, Nov 2025. [Online].
		Available: \url{https://doi.org/10.1088/2058-9565/ae1e98}
		\BIBentrySTDinterwordspacing

        \bibitem{Du2026}
		\BIBentryALTinterwordspacing
       Q.~Du, J.~Xu, Y.~Jin, W.~Wang, Z.~Tu, Y.~Liu, H.~Lian, Y.~Zhu, Z.~Shan, ``From Noise Modeling to Layout Optimization: A Framework for Quantum Circuit Fidelity Enhancement With Machine Learning,''
		\emph{Quantum Sci. Technol.}, vol.~9, p. 3, Mar 2026. [Online].
		Available: \url{https://doi.org/10.1002/qute.202500464}
		\BIBentrySTDinterwordspacing 

        \bibitem{Ghosh2012}
		\BIBentryALTinterwordspacing
       J.~Ghosh, A.G.~Fowler, M.~Geller, R.~Michael, ``Surface code with decoherence: An analysis of three superconducting architectures,''
		\emph{Phys. Rev. A}, vol.~86, no. 6, p. 062318, Dec 2012. [Online].
		Available: \url{https://link.aps.org/doi/10.1103/PhysRevA.86.062318}
		\BIBentrySTDinterwordspacing

        \bibitem{Cai2023}
		\BIBentryALTinterwordspacing
       Z.~Cai, R.~Babbush, S.C.~Benjamin, S.~Endo, W.J..~Huggins, Y.~ Li, J.R.~McClean, T.E.~O'Brien, ``Quantum error mitigation,''
		\emph{Rev. Mod. Phys.}, vol.~95, no. 4, p. 045005, Dec 2023. [Online].
		Available: \url{https://link.aps.org/doi/10.1103/RevModPhys.95.045005}
		\BIBentrySTDinterwordspacing

        \bibitem{ahmad2024_mit}
		\BIBentryALTinterwordspacing
		H.~G. Ahmad, R.~Schiattarella, P.~Mastrovito, A.~Chiatto, A.~Levochkina,
		M.~Esposito, D.~Montemurro, G.~P. Pepe, A.~Bruno, F.~Tafuri, A.~Vitiello,
		G.~Acampora, and D.~Massarotti, ``Mitigating errors on superconducting
		quantum processors through fuzzy clustering,'' \emph{Advanced Quantum
			Technologies}, vol.~7, no.~7, p. 2300400, 2024. [Online]. Available:
		\url{https://advanced.onlinelibrary.wiley.com/doi/abs/10.1002/qute.202300400}
		\BIBentrySTDinterwordspacing

        \bibitem{Viola1998}
		\BIBentryALTinterwordspacing
		L.~Viola, S.~Loyd, ``Dynamical suppression of decoherence in two-state quantum systems,'' \emph{Phys. Rev. A}, vol.~58, no.~4, p. 2733-2744, Oct 1998. [Online]. Available:
		\url{https://link.aps.org/doi/10.1103/PhysRevA.58.2733}
		\BIBentrySTDinterwordspacing

        \bibitem{levochkina2024investigating}
		A.~Y. Levochkina, H.~G. Ahmad, P.~Mastrovito, I.~Chatterjee, G.~Serpico,
		L.~Di~Palma, R.~Ferroiuolo, R.~Satariano, P.~Darvehi, A.~Ranadive
		\emph{et~al.}, ``Investigating pump harmonics generation in a snail-based
		traveling wave parametric amplifier,'' \emph{Superconductor Science and
			Technology}, vol.~37, no.~11, p. 115021, 2024.

        \bibitem{Schulze2021}
		\BIBentryALTinterwordspacing
		L.~Schulze, J.-R.~Lahmann, ``Evaluating error mitigation strategies for entangled quantum states on near-term quantum computers,'' \emph{INFORMATIK 2021}, p. 943-960, 2021. [Online]. Available:
		\url{https://doi/10.18420/informatik2021-079}
		\BIBentrySTDinterwordspacing
		
\end{thebibliography}



\section*{Supplementary material}

# Supplementary Materials for Quantum Circuit-Based Adaptation for Credit Risk Analysis

## VII. Conditional default probability as quantum state probability: a mathematical demonstration

### A. Gaussian Conditional Independence Model

In the Gaussian conditional-independence model [1], [2], the default probability at the obligor-level conditioned on a latent Gaussian factor $z$ is

$$PD_k(z) = \Phi\left(\frac{\Phi^{-1}(p_k^0) - \sqrt{\rho_k}\, z}{\sqrt{1 - \rho_k}}\right), \qquad (17)$$

with the default baseline level $p_k^0$, the correlation of assets $\rho_k$ and $\Phi$ the standard normal Cumulative Distribution Function (CDF). Our circuit is designed so that the probability of measuring $|1\rangle$ on a target qubit matches this conditional default probability. We write

$$PD_k(z) \equiv \Pr(|1\rangle) = \sin^2(\alpha_k z + \beta_k), \qquad (18)$$

which is equivalent to linearizing

$$g(z) := \arcsin\sqrt{PD_k(z)} \approx \alpha_k z + \beta_k \quad \text{around } z = 0. \quad (19)$$

Let $f$ denote standard normal PDF, and define

$$\psi_k := \frac{\Phi^{-1}(p_k^0)}{\sqrt{1 - \rho_k}}. \qquad (20)$$

The offset is the zeroth-order term of $g$:

$$\beta_k = g(0) = \arcsin\sqrt{\Phi(\psi_k)}. \qquad (21)$$

By the chain rule, the slope is

$$\alpha_k = \frac{dg}{dz}\bigg|_{z=0} = -\frac{1}{\sqrt{1 - \Phi(\psi_k)}} \cdot \frac{1}{2\sqrt{\Phi(\psi_k)}} \cdot f(\psi_k) \cdot \frac{\sqrt{\rho_k}}{\sqrt{1 - \rho_k}}. \qquad (22)$$

Hence the first-order approximation reads

$$\arcsin\sqrt{PD_k(z)} \approx \alpha_k z + \beta_k, \qquad PD_k(z) \approx \sin^2(\alpha_k z + \beta_k), \qquad (23)$$

so the required single-qubit gate is a linear-rotation block

$$R_y\big(2(\alpha_k z + \beta_k)\big). \qquad (24)$$

When the latent factor is represented on an $n$-qubit register discretizing $[-z_{\max}, z_{\max}]$ with grid step

$$\Delta z = \frac{2 z_{\max}}{2^n - 1}, \qquad (25)$$

the circuit angles are rescaled to the coded grid:

$$\tilde{\alpha}_k = \Delta z\, \alpha_k = \frac{2 z_{\max}}{2^n - 1}\, \alpha_k, \qquad \tilde{\beta}_k = \beta_k - \alpha_k z_{\max}, \qquad (26)$$

and the implemented gate becomes

$$R_y\big(2(\tilde{\alpha}_k\, \hat{z}_{\text{code}} + \tilde{\beta}_k)\big), \qquad (27)$$

where $\hat{z}_{\text{code}} \in \{0, 1, \ldots, 2^n - 1\}$ denotes the integer basis value encoded by the $n$-qubit $z$-register.

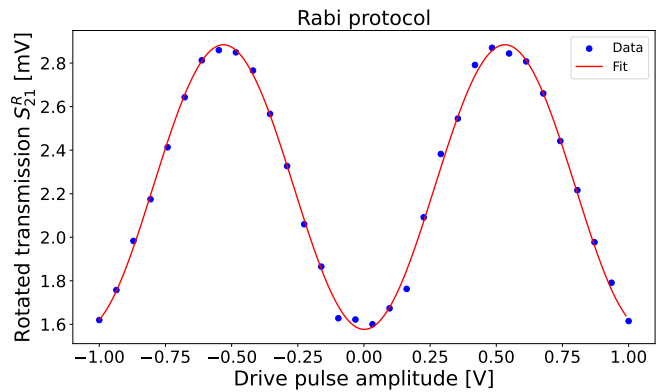

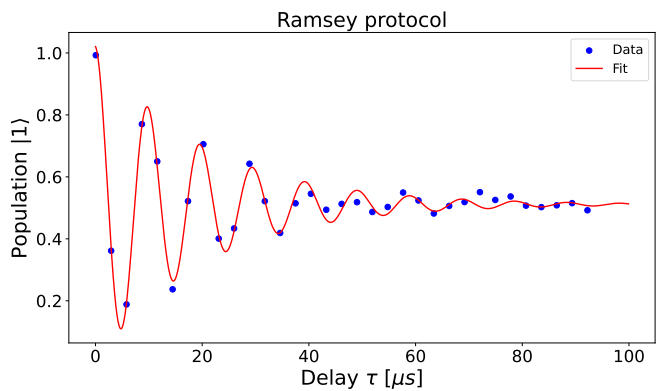

Fig. S1: In **(a)**: example of a Rabi oscillation measurement; in **(b)**: example of a Ramsey Interferometry experiment.

## VIII. Calibration, Fidelity and Readout

In this section, we describe the main protocols used for calibration and optimization of single-qubit X and Y gates, how to benchmark single-qubit gate fidelities and report on the acquired experimental data obtained before running the experiments discussed in the main manuscript.

### A. Drive signal calibration protocols

The first step of the drive signals' calibration process consists in time-domain measurements. The Rabi oscillations measurements provide the estimation of the $\pi$-pulse amplitude, i.e. the microwave drive signal amplitude necessary to bring the qubit from the ground state $|0\rangle$ to the first excited state $|1\rangle$ for a fixed drive pulse duration. The operative procedure to measure the Rabi oscillations consists in sending a drive signal followed by a readout pulse, by changing the amplitude of the drive tone.

Ramsey interferometry protocol allows to tune the drive signal frequency by preparing the qubit on the Bloch sphere equator applying a $X_{\frac{\pi}{2}}$-pulse (i.e. halving the $\pi$-pulse amplitude measured throuh Rabi oscillations) and then leaving the qubit free to evolve spontaneously for a time $\tau$ before that a second $X_{\frac{\pi}{2}}$-pulse projects the Bloch vector back to the z-axis. When inferring the qubit state, in presence of

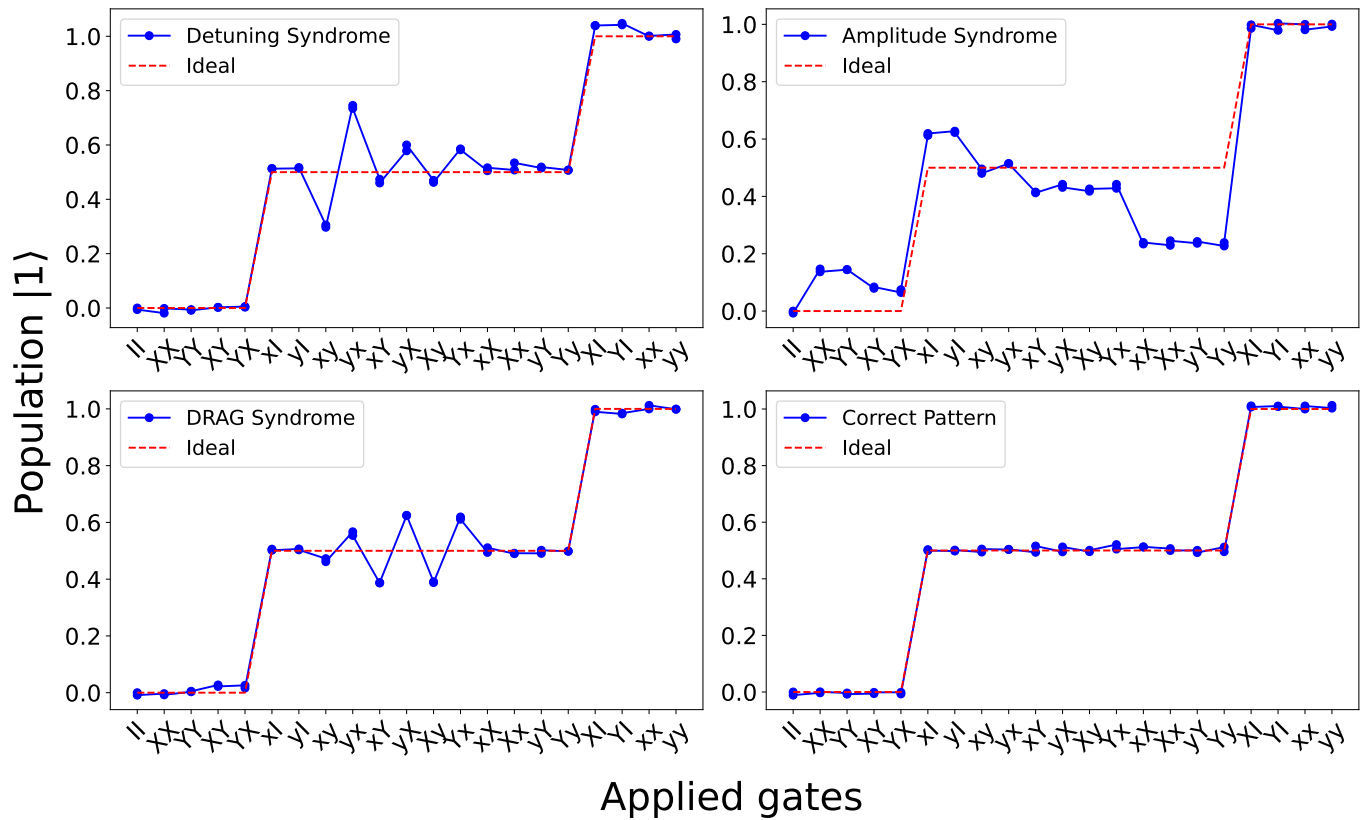

Fig. S2: Pattern signatures for of power amplitude, detuning and DRAG syndromes, and corrected pattern, showing the ideal $z$ projection as a function of *AllXY* sequences.

detuning between the drive and qubit frequencies, i.e. when the drive signal is off-resonance with the qubit frequency, damped oscillations with an exponential decay arise. On the contrary, the output features only an exponential decay.

### B. Detecting syndromes for control optimization: All-XY sequence

Higher quality rotations are desirable for applications where a high fidelity is needed, as for quantum algorithms implementation. The *AllXY* sequence tests the result of all combinations of one or two single-qubit gates of the X and Y type being sensitive to a variety of error syndromes [3]. During the protocol, 21 different pulse combinations, belonging to the "Clifford set" of quantum gates [4], are performed on a qubit, producing rotations around the $x$- and $y$-axes by an angle of $\pi/2$ or $\pi$ before measuring its state. Each pulse combination is sensitive to various errors, which will have a distinct fingerprint in the deviation from the ideal response (Fig. S2), providing a quick way of diagnosing syndromes [3]. It is possible to detect mainly three types of syndromes, which can be tackled with a specific correction protocol:

- Power syndrome: related to drive signal amplitude, it yelds a characteristic "step" pattern, shown in Fig. S2, due to over- or under-rotations of the gates. To correct amplitude errors, in this work we used the **Flipping** (or error amplification) protocol. This protocol consists in preparing the qubit on the equator of the Bloch sphere

by applying a $\pi/2$-pulse. Then, $2N$ $\pi$-pulses are applied, with an increasing $N$ (Fig. S3 (a)). Each additional $\pi$-pulse amplifies the power error syndrome, causing the qubit state to oscillate as a function of $N$. By fitting these oscillations, it is possible to identify the control pulse amplitude to correct this behavior, and observe instead a linear dependence of the qubit state population (Fig. S3 (b)).

- Detuning syndrome: caused by the detuning between drive signal frequency and qubit resonance frequency, it yelds a characteristic "zig-zag" pattern on the $x$ and $y$ rotations. Therefore, this syndrome can be corrected by implementing Ramsey protocol to fit the detuning and estimate the correction needed to set right drive signal frequency.

- DRAG syndrome: related to phase errors, it's mainly due the IQ mixer imperfections, such as amplitude imbalance and skewness. The DRAG technique corrects both gate phase errors and leakage caused by the presence of higher excited states of weakly anharmonic transmon qubit [5]. Through **Motzoi** protocol, it is possible to efficiently tune-up the *DRAG* parameter [6] as follows: by taking two of the *AllXY* pulses which exhibit the opposite sign of error (for instance $Y_\pi X_{\frac{\pi}{2}}$ and $X_\pi Y_{\frac{\pi}{2}}$) and performing them as a function of the parameter proportional to the *DRAG* derivative correction, one obtains two lines that cross at the point where the *DRAG* optimal coefficient is

optimal (Fig. S4).

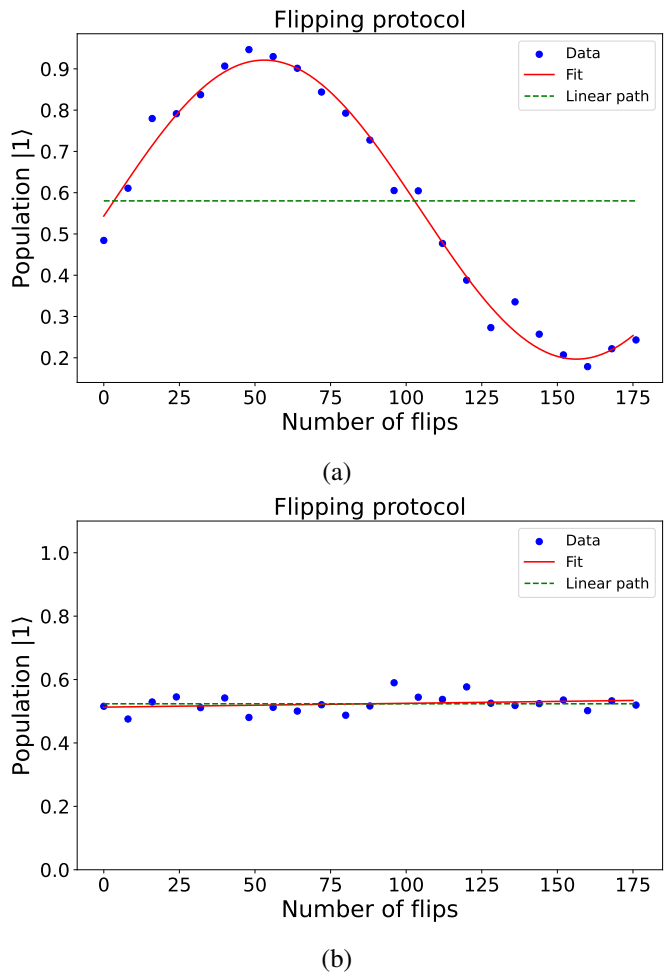

(a)

(b)

Fig. S3: Flipping protocol before (**a**) and after (**b**) applying power error syndrome correction.

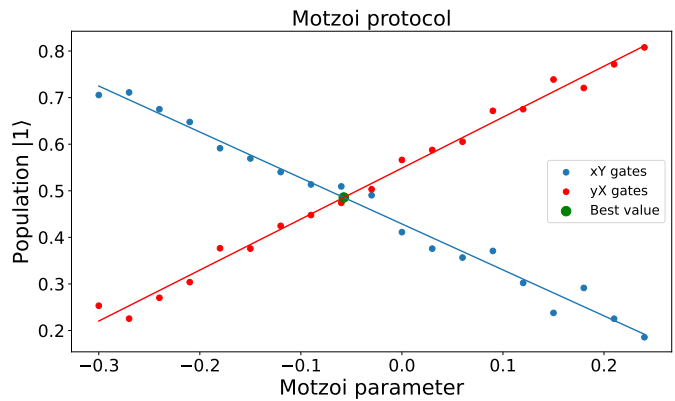

Fig. S4: Motzoi protocol is performed by choosing two of the *AllXY* pulse sequences with opposite sign errors, and measure them both with varying *DRAG* coefficient: the optimal DRAG value is located at the crossing point of two resulting lines [3].

### C. Gate fidelity

In the main manuscript, we used the Randomized Benchmarking (RB) and the Interleaved Randomized Benchmarking (IRB) protocol in order to assess the average single-qubit gate fidelity and the CZ two-qubit gate fidelity, respectively [7].

The Randomized Benchmarking protocol performs an average gate fidelity evaluation that is robust to State Preparation and Measurements (SPAM) errors, gaining information on the hardware gate implementation [8]. RB protocol consists of applying a random sequence of $N$ gates, randomly picked from the Clifford gate set $\{C_i\}$, followed by their inverse Clifford gate before measuring the qubit state population. Ideally, the measured state should be identical to the initial state, i.e. ground state $|0\rangle$, but the measured state will differ due to the errors induced by the gate implementation. The number of gates in a sequence is chosen to be $N_{Cl} = [1, 2, 4, 8, 16, 32, 64, 128, 256, 512, ...]$ to easily obtain an exponential decay trend (Fig. S5). To perform repeated measurements with different gate combinations, for each sequence $N_{Cl}$, random gates are sorted through a certain number of seeds, i.e. random number generators. Therefore, the survival probability $F_{|0\rangle}$, i.e. the average sequence gate fidelity is measured for each seed. After the measurement, the depolarizing error $p$ is fitted using the model

$$F_{|0\rangle} = Ap^{N_{Cl}} + B. \tag{28}$$

Here $A$ and $B$ parameters give information about SPAM errors. Then, the gate fidelity over all the Clifford sequences is given by [8]:

$$F_{Cl} = 1 - \frac{d-1}{d}(1-p). \tag{29}$$

Lastly, it is possible to get an estimation of average gate fidelity per applied gate as [8]:

$$F_{avg.gate} = (F_{Cl})^{\frac{1}{1.875}}, \tag{30}$$

where the exponent of the root-square is related to the decomposition of each gate into the standard minimal sequence of $\pi$- and $\pi/2$-pulses around the $x$ and $y$ axes, requiring an average $\langle N_P \rangle = 1.875$ pulses per Clifford. In Fig.S5, we show the RB experiment performed simultaneously on the three qubits used for the implementation of the quantum circuits in the main manuscript.

While the RB allows for estimating the average single-qubit gate fidelity, the Interleaved Randomized Benchmarking (IRB) allows for estimating the average error of an individual quantum gate. Therefore, the CZ gate (the native two-qubit gate in our processor) fidelity can be estimated by interleaving the CZ gate into the standard RB protocol [9]. First, a standard RB experiment on the two-qubit allows to recover the depolarizing error $p_0$ without the insertion of the CZ gate in the sequence. Then, the RB is repeated by including the CZ after each Clifford gate randomly applied. As for the standard RB, the depolarizing error $p_1$ can be fitted by the exponential decay. Finally, the average fidelity for the CZ gate can be estimated by:

$$F_{cz,naive} = (F_{Cl})^{\frac{1}{1.5}}. \tag{31}$$

In Fig.S6, we show the IRB experiment for the two pairs investigated in the main manuscript.

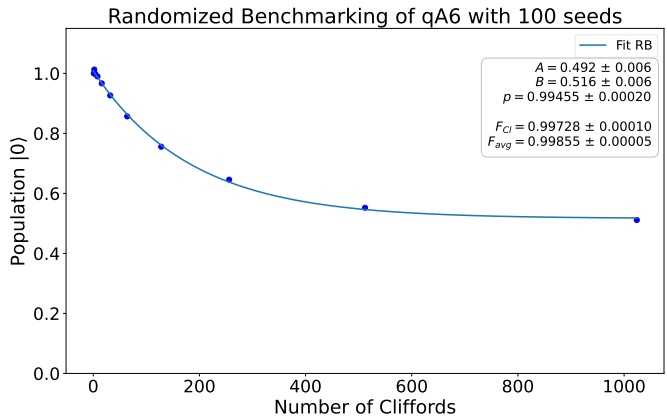

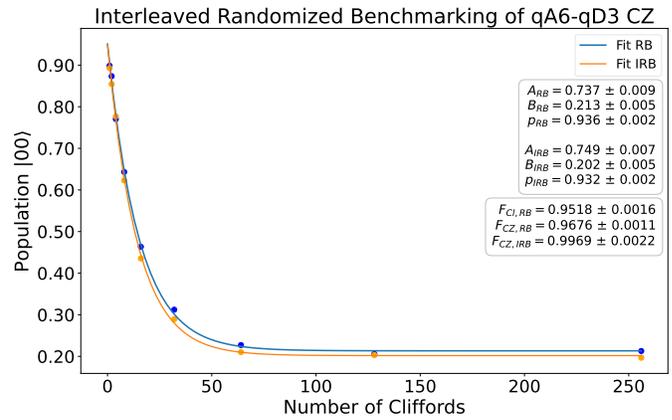

(a)

(a)

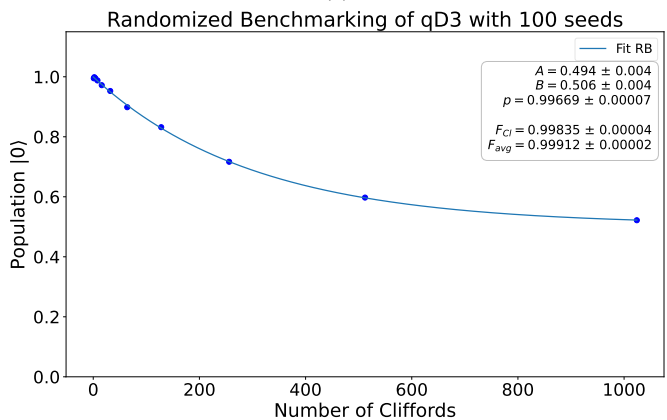

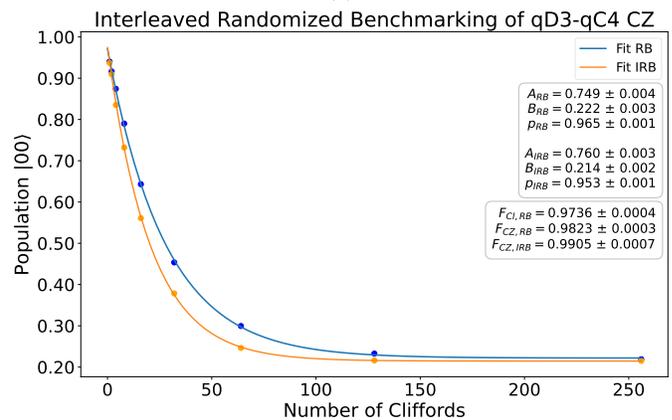

(b)

(b)

Fig. S6: Interleaved Randomized Benchmarking protocol for CZ gate implemented on qubit couples A6-D3, and D3-C4.

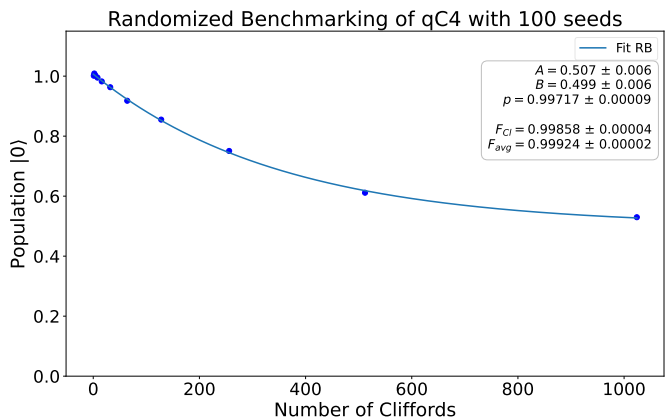

(c)

Fig. S5: Simultaneous Randomized Benchmarking protocol with 100 seeds of qubits A6, D3, C4.

### D. Readout

A fundamental requirement for the implementation of quantum algorithms is the ability to readout the state of a quantum register in such a way to prevent the quantum state collapse, i. e. within a quantum non-demolitive readout. Transmon qubits allows to do it by dispersively coupling each qubit with a superconducting readout resonator. As detailed in Ref. [12], the assignment state probability of a qubit is obtained by measuring multiple times the signal across the

resonator (Single-Shot Readout - SSRO) and counting how many times the qubit has been in the computational basis state $\{|0\rangle, |1\rangle\}$. This eventually requires to identify a discrimination threshold, which is obtained through an initialization experiment: the qubit is prepared in the two computational basis states, SSRO is performed and the discrimination threshold is identified. The procedure can be extended to more qubits, and the counting mechanism used in this work exploits the conditional readout proposed in Ref. [12]. As a consequence, the ability to efficiently readout the output of a quantum circuit strongly depends on the readout parameters that allows in turn to efficiently discriminate the state. Among them, the microwave excitation tone sent to the resonator, as well as the acquisition signal used to measure the complex voltage across the resonator, in which the qubit state is encoded, have been calibrated by performing repeated SSRO initialization experiments as a function of their pulse parameters (duration, amplitude, frequency): the best parameters have been chosen as the ones for which both leakage towards non-prepared states and readout fidelity was maximized.

Finally, before every quantum circuit we implement the SSRO initialization experiment to identify the discrimination threshold right before any quantum circuit experiment. In Fig.S7, for example, we show the initialization matrix for the

GCI algorithm experiment.

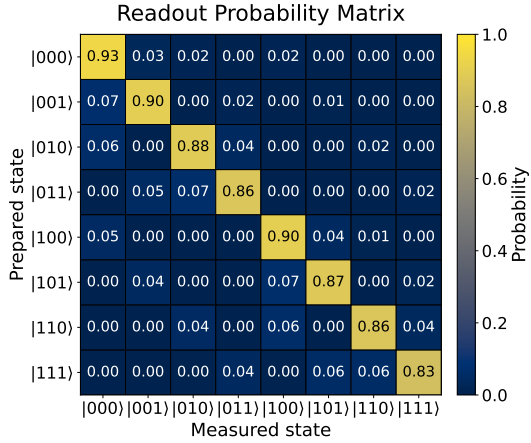

Fig. S7: Single Shot Readout Probability Matrix for the three qubit register D3-A6-C4.

## IX. DETAILS ON THE HARDWARE-AWARE VARIATIONAL QUANTUM ALGORITHMS

### A. Two-qubit Gaussian Loading: CNOT implementation

As shown in Section C. of the main manuscript, we investigated the role of the hardware connectivity in the implementation of the quantum algorithm by performing two-qubit Gaussian distribution generation on two different two-qubit registers.

For A6 and D3, we found an optimal combination angle $\theta_0 = 90°$ and $\theta_1 = 222°$. We also tested the CNOT directionality effect by changing the role of control and target qubit. We recall that the $CNOT$ gate is implemented as $CNOT_{0,1} = (\mathbb{I} \otimes H)CZ(\mathbb{I} \otimes H)$ (Fig. S8). Specifically, the implementation of the CZ gate consists of moving the highest-frequency qubit (i.e. the tuned qubit) from the sweet-spot towards the lowest-frequency qubit (spectator qubit). Here, for example, we used a Sudden-Net-Zero Flux pulse, opportunely calibrated. When the tuned qubit corresponds to the control qubit, it will only relive a flux-pulse at the hardware level, since Hadamard gate will be applied on the spectator qubit. If, instead, it is the target qubit, the CZ gate will follow a Hadamard gate. However, although the CZ gate should only

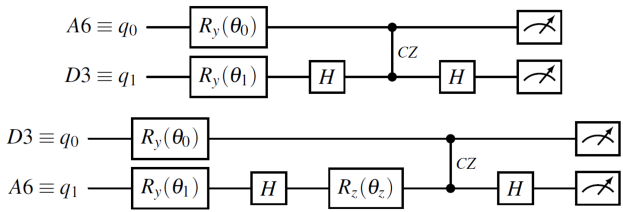

Fig. S8: Two-qubit quantum circuit for the Gaussian loading on A6-D3 register: (top), the implementation of $CNOT_{A6,D3}$, (bottom) the implementation of $CNOT_{D3,A6}$ with counter phase $R_z(\theta)$ gate, i.e. when the tuned qubit is the targeted one.

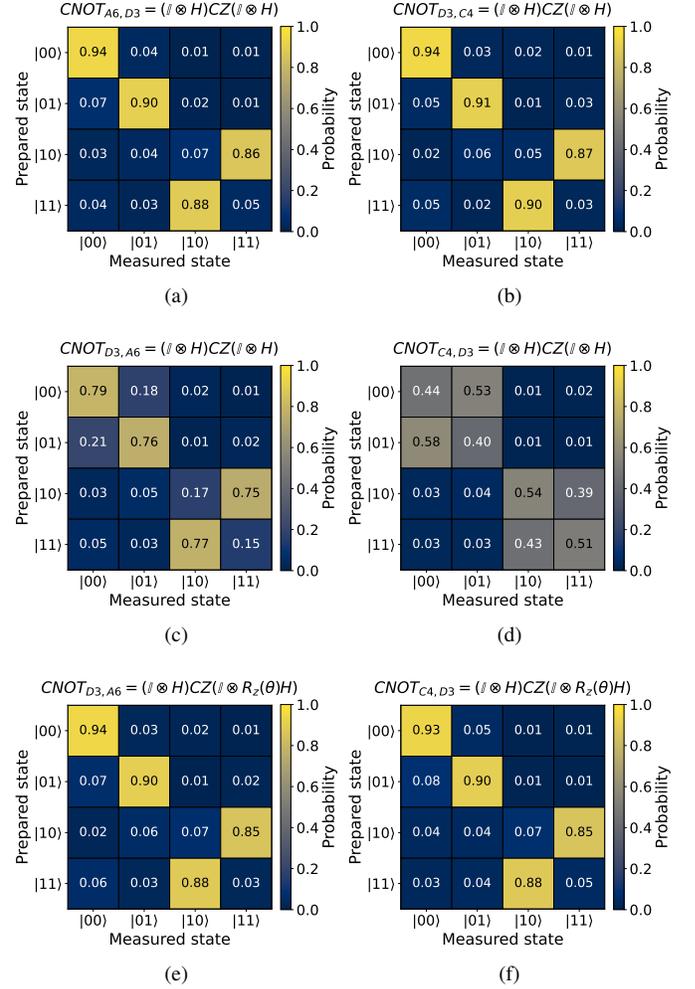

Fig. S9: In **(a)** and **(b)** $CNOT_{A6,D3}$ and $CNOT_{D3,C4}$, respectively. In **(c)** and **(d)** $CNOT_{D3,A6}$ and $CNOT_{C4,D3}$, respectively, i.e. when the tuned qubit is the target of the $CNOT$ gate. In **(e)** and **(f)** $CNOT_{D3,A6}$ and $CNOT_{C4,D3}$ with the counter gate $Rz(\theta)$, with $\theta = -135°$ for D3-A6 register and $\theta = -90°$ for C4-D3 register.

adiabatically move the qubit from the sweetspot, in reality this dynamics may cause the qubit to gain an additional single-qubit phase, compromising the quality of the CNOT. In order to counteract this effect, an additional $R_z(\theta_z)$ gate is added between the first Hadamard gate and the CZ, acting as a counter phase gate. By repeatedly measuring the Truth table of the CNOT as a function of the counter-phase gate angle, we have identified $\theta = -135°$ as the best correction angle for D3-A6 couple and $\theta = -90°$ for C4-D3 couple. The results are reported in Fig.S9.

Finally, following this argument, we asked ourselves whether the optimal angles for generating two-qubit Gaussian distributions would depend on the roles of the target and control in the register. Therefore, we performed the algorithm depicted in Fig.S8, changing the role of target and control in the pair. We obtained a similar optimal angle combination, i.e. $\theta_0 = 90°$ and $\theta_1 = 224°$, meaning that the optimal

**TABLE S1:** Two-qubit circuit output on D3-C4 register as a function of $\theta_1$ and fixed $\theta_0 = 90°$ by preparing the $|11\rangle$ state. For $\theta_1 \in \{90°, 270°\}$ probability amplitudes feature an upside down gaussian-like shape, while for $\theta_1 \in \{270°, 450°\}$ feature a normal gaussian-like shape.

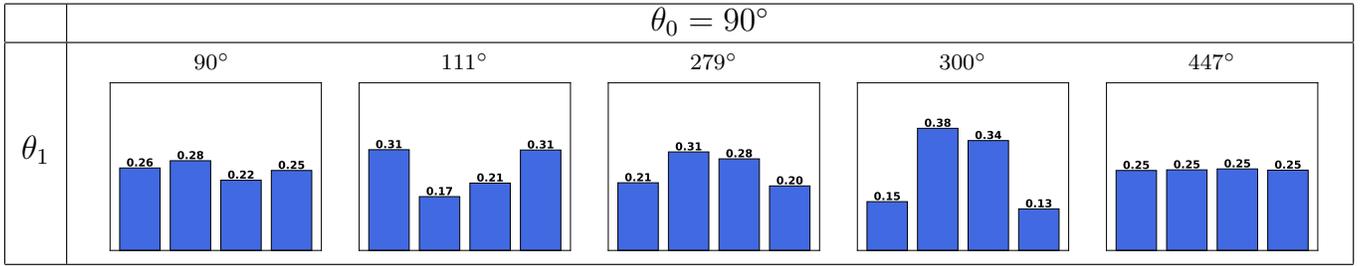

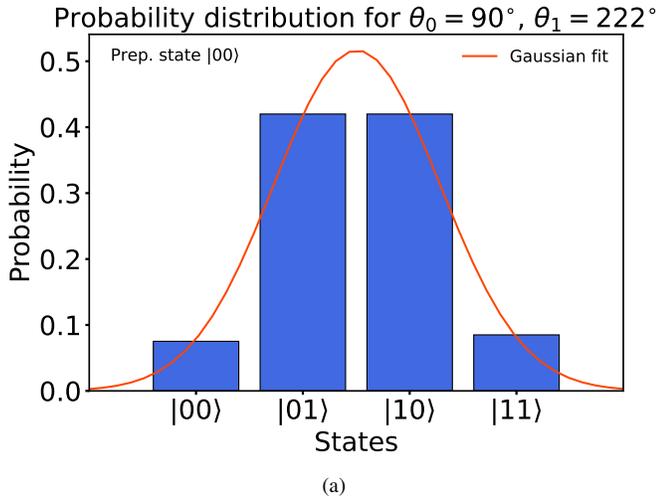

(a)

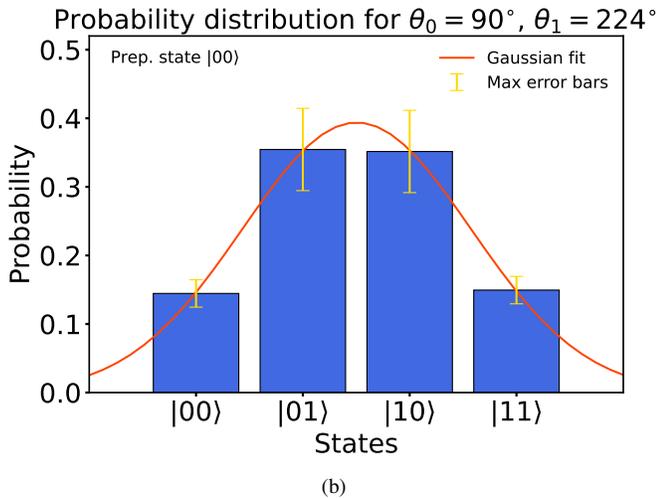

(b)

Fig. S10: Two-qubit Gaussian generation for the A6-D3 register: in **(a)** D3 plays the role of target qubit, while in **(b)** A6 is the targeted one.

angles don't depend on the role of the qubits, but only on the exploited two-qubit register (Fig.S10).

### B. Alternative state preparation

The goal of the work is to load a two-qubit normal distribution in the GCI quantum circuit. In Tab.S1, we show the same experiment described in Sec. A "Two-qubit normal distribution

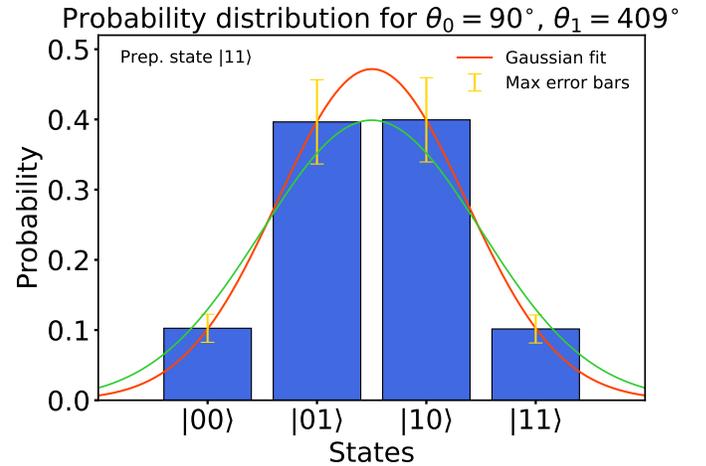

Fig. S11: Two-qubit Gaussian distribution at the optimal $\theta_0$ and $\theta_1$ angle and corresponding Gaussian fit for D3-C4 pair prepared in $|11\rangle$ state. In green the N(0,1) distribution.

encoding and loading", by preparing the D3-C4 qubit pair in $|11\rangle$ initial state, rather than in the ground state. In this case, the general trend is opposite to what we observe for the initial state $|00\rangle$. We followed the same variational approach discussed in the main text to obtain a standard normal distribution, and in Fig.S11 the gaussian distribution is obtained with the optimal angle configuration $\theta_0 = 90°$ and $\theta_1 = 409°$.

### C. Three-qubit Gaussian Loading: variational experiments.

As done for the two-qubit experiment, in Fig. S12 we show the experimental probability amplitudes for the three-qubit basis states as a function of the angle $\theta_1$ and $\theta_2$: they show a reasonable agreement with the theoretical expectation. Remarkably, as discussed in the main text, we can revover a uniform distribution for different angles configurations.

### D. Error bars

In order to estimate the error bars on the probability amplitudes, we performed a statistics of 100 measurements for the optimal set of angles. Here we show an example for $\theta_0 = 90°$ and $\theta_1 = 191°$ for the two-qubit register D3-C4. In particular, we assumed the deviation from the symmetric case, i.e. $P_{|00\rangle} = P_{|11\rangle}$ and $P_{|01\rangle} = P_{|10\rangle}$, so

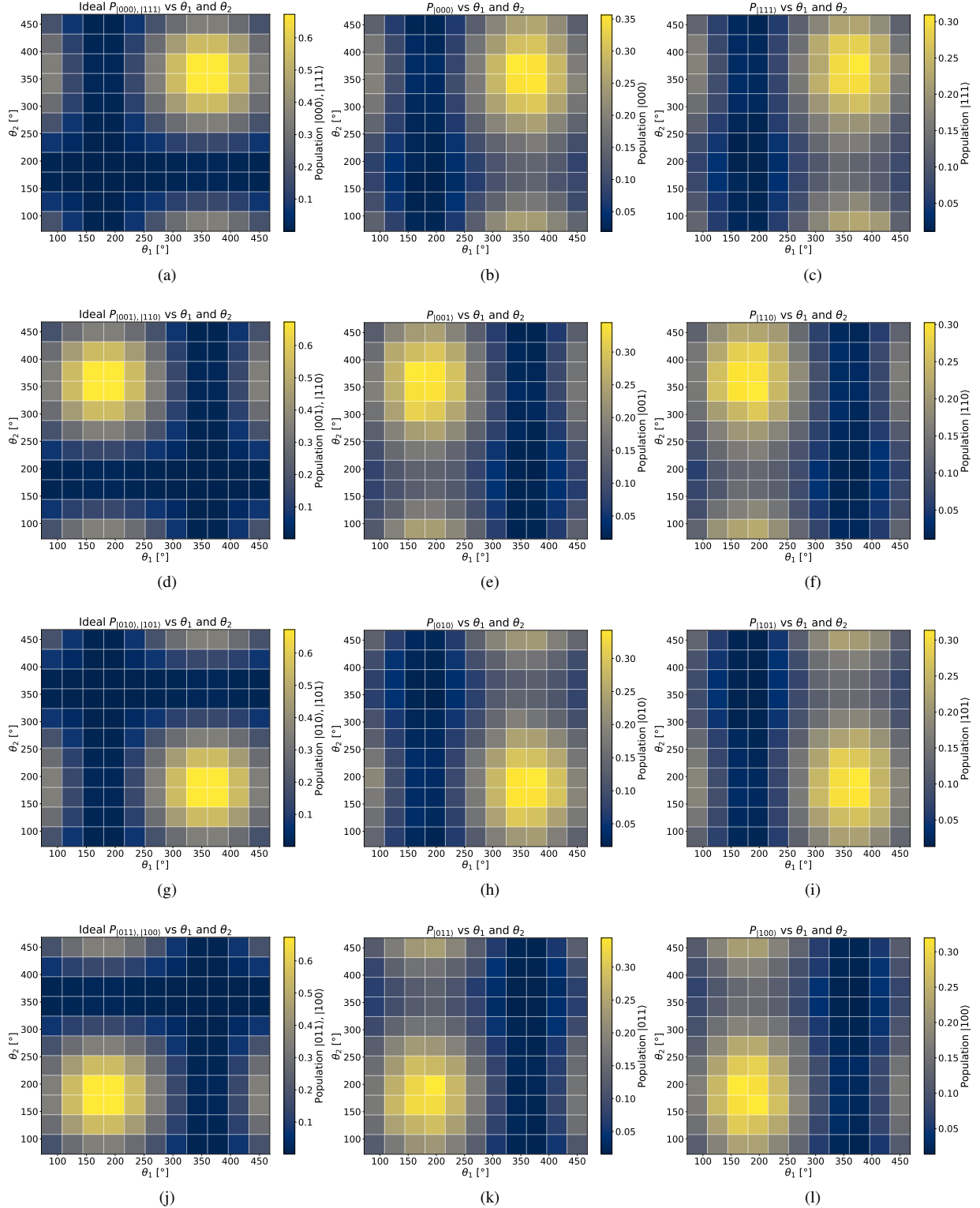

Fig. S12: Comparison between probability amplitudes of $|000\rangle, |001\rangle, |010\rangle, |011\rangle, |100\rangle, |101\rangle, |110\rangle, |111\rangle$ states and theoretical model as a function of $\theta_1$ and $\theta_2$ with fixed $\theta_0 = 90°$

$\Delta P_{|00\rangle,|11\rangle} = P_{|00\rangle} - P_{|11\rangle}$ and $\Delta P_{|01\rangle,|10\rangle} = P_{|01\rangle} - P_{|10\rangle}$. We show in Fig.S13 that the mean values are $\Delta P_{|00\rangle,|11\rangle} = 0.02$ and $\Delta P_{|01\rangle,|10\rangle} = 0.06$.

We proceeded in the same way for the three-qubit gaussian

generation experiment by performing a 100 measurements statistic with the optimal angles combination for D3-A6-C4 register: $\theta_0 = 90°$, $\theta_1 = 212.5°$ and $\theta_2 = 104.5°$ (Fig. S14). The mean values are $\Delta P_{|000\rangle,|111\rangle} = 0.01$, $\Delta P_{|001\rangle,|110\rangle} =$

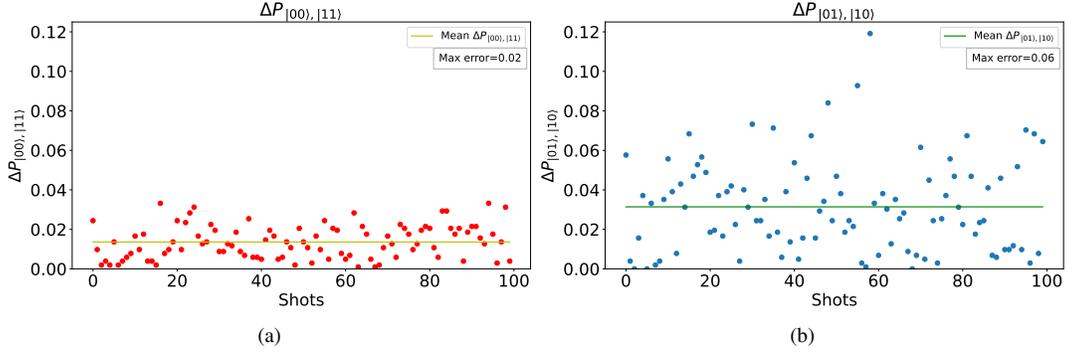

Fig. S13: Error bars estimation from 100 measurement statistic for the D3-C4 register.

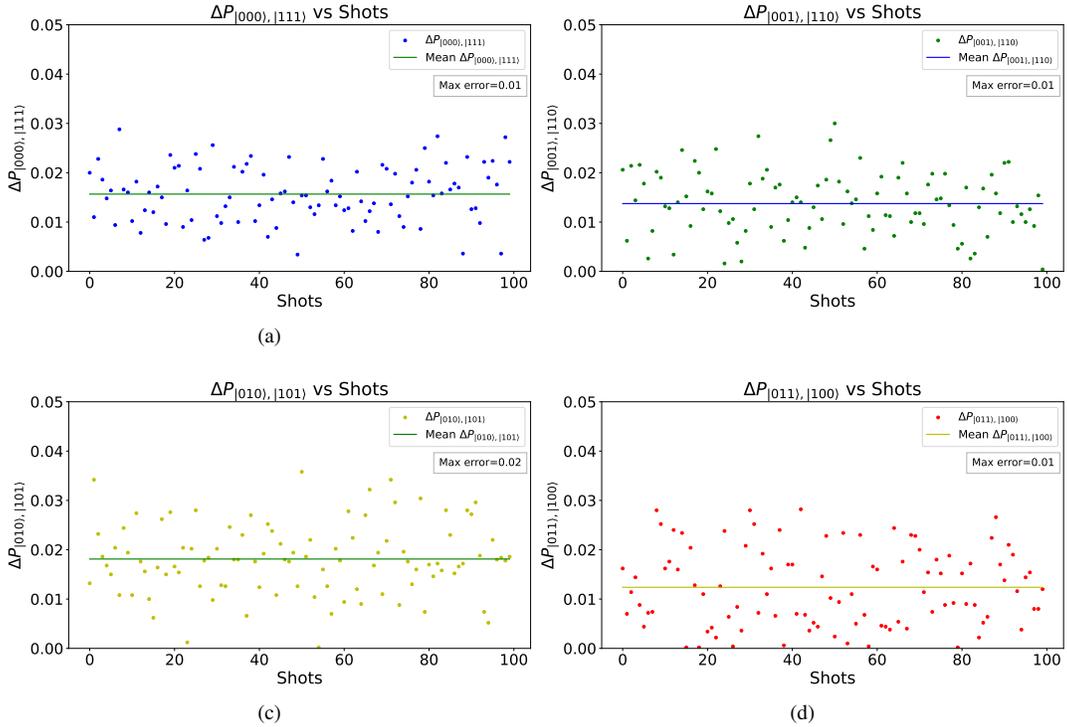

Fig. S14: Error bars estimation from 100 measurement statistic for the D3-A6-C4 register.

$0.01$, $\Delta P_{|010\rangle,|101\rangle} = 0.02$ and $\Delta P_{|011\rangle,|100\rangle} = 0.01$.

### E. Use-case GCI algorithm: experiments as a function of the asset rotation angle

In Tab.S2, we show the probability amplitudes as a function of $R_y(\theta_2)$, i.e. the asset preparation gate, at fixed $\theta_3 = 90°$ and $\theta_4 = 180°$ (Fig. S15).

## REFERENCES

[1] Q. Community, "Credit risk analysis," https://qiskit-community.github.io/qiskit-finance/tutorials/09_credit_risk_analysis.html, qiskit Finance 0.4.1. Accessed: 2025-09-24.

[2] M. RUTKOWSKI and S. TARCA, "Regulatory capital modeling for credit risk," *International Journal of Theoretical and Applied Finance*, vol. 18, no. 05, p. 1550034, 2015. [Online]. Available: https://doi.org/10.1142/S021902491550034X

[3] M. Reed, "Entanglement and quantum error correction with superconducting qubits," 2013. [Online]. Available: https://arxiv.org/abs/1311.6759

[4] S. Bravyi and A. Kitaev, "Universal quantum computation with ideal clifford gates and noisy ancillas," *Phys. Rev. A*, vol. 71, p. 022316, Feb 2005. [Online]. Available: https://link.aps.org/doi/10.1103/PhysRevA.71.022316

[5] J. M. Chow, L. DiCarlo, J. M. Gambetta, F. Motzoi, L. Frunzio, S. M. Girvin, and R. J. Schoelkopf, "Optimized driving of superconducting artificial atoms for improved single-qubit gates," *Phys. Rev. A*, vol. 82, p. 040305, Oct 2010. [Online]. Available: https://link.aps.org/doi/10.1103/PhysRevA.82.040305

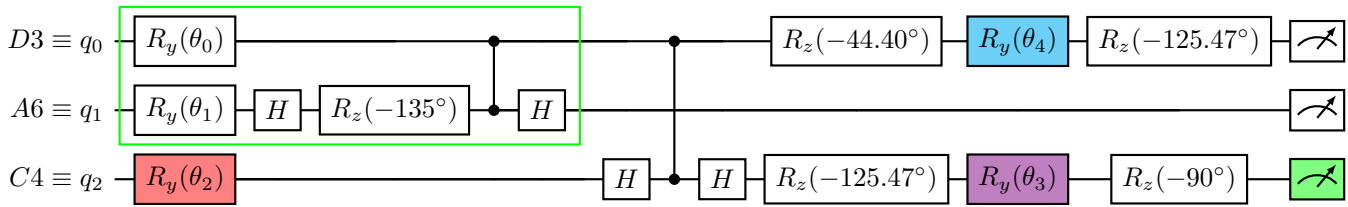

Fig. S15: Use-Case quantum circuit.

**TABLE S2:** GCI circuit output as a function of rotation asset $\theta_2$ at fixed $\theta_3 = 90°$ and $\theta_4 = 180°$.

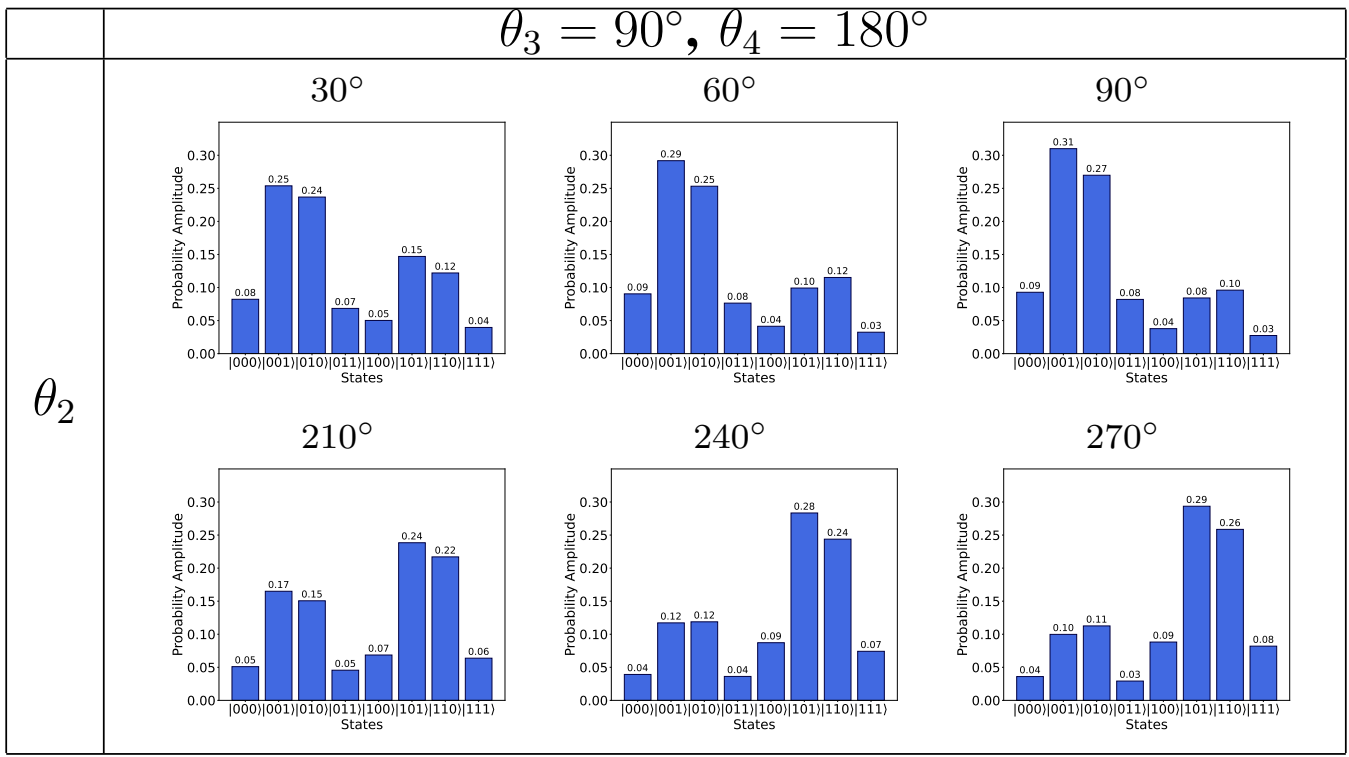

[6] F. Motzoi, J. M. Gambetta, P. Rebentrost, and F. K. Wilhelm, "Simple pulses for elimination of leakage in weakly nonlinear qubits," *Phys. Rev. Lett.*, vol. 103, p. 110501, Sep 2009. [Online]. Available: https://link.aps.org/doi/10.1103/PhysRevLett.103.110501

[7] A. Hashim, L. B. Nguyen, N. Goss, B. Marinelli, R. K. Naik, T. Chistolini, J. Hines, J. Marceaux, Y. Kim, P. Gokhale, T. Tomesh, S. Chen, L. Jiang, S. Ferracin, K. Rudinger, T. Proctor, K. C. Young, I. Siddiqi, and R. Blume-Kohout, "Practical introduction to benchmarking and characterization of quantum computers," *PRX Quantum*, vol. 6, p. 030202, Aug 2025. [Online]. Available: https://link.aps.org/doi/10.1103/PRXQuantum.6.030202

[8] E. Magesan, J. M. Gambetta, and J. Emerson, "Scalable and robust randomized benchmarking of quantum processes," *Phys. Rev. Lett.*, vol. 106, p. 180504, May 2011. [Online]. Available: https://link.aps.org/doi/10.1103/PhysRevLett.106.180504

[9] E. Magesan, J. M. Gambetta, B. R. Johnson, C. A. Ryan, J. M. Chow, S. T. Merkel, M. P. da Silva, G. A. Keefe, M. B. Rothwell, T. A. Ohki, M. B. Ketchen, and M. Steffen, "Efficient measurement of quantum gate error by interleaved randomized benchmarking," *Phys. Rev. Lett.*, vol. 109, p. 080505, Aug 2012. [Online]. Available: https://link.aps.org/doi/10.1103/PhysRevLett.109.080505



\end{document}